\documentclass[aps,prd,twocolumn,superscriptaddress,groupedaddress]{revtex4}
\usepackage[T1]{fontenc}
\usepackage[english]{babel}
\usepackage{amssymb}
\usepackage{cancel}
\usepackage{color,graphicx}
\usepackage{amsmath}
\usepackage{amsbsy}
\usepackage{amsthm}
\usepackage{bbm}
\usepackage{booktabs}
\usepackage{epsfig}
\usepackage{float}
\usepackage{graphicx}
\usepackage{dcolumn}
\usepackage{bbm}
\usepackage{color,epstopdf}
\usepackage{amscd}
\usepackage{amsfonts}  
\usepackage[]{amsmath}
\usepackage{amssymb}    
\usepackage{mathrsfs}
\usepackage{verbatim}
\usepackage[]{cases}
\usepackage{amsmath}
\usepackage{wasysym}
\usepackage[utf8]{inputenc}
\usepackage{mathtools}

\usepackage{bbold}
\usepackage{braket}
\usepackage{graphicx}
\usepackage{todonotes}
\usepackage{cleveref}

\begin{document}
\title{QTAM: Q-Transform Amplitude Modulation}
\author{L. Asprea}
\email{lorenzo.asprea@to.infn.it}
\affiliation{Istituto Nazionale di Fisica Nucleare, Via Giuria 1, Torino, Italy}      
\author{F. Sarandrea}
\affiliation{Istituto Nazionale di Fisica Nucleare, Via Giuria 1, Torino, Italy}  
\author{A. Romano}
\affiliation{Istituto Nazionale di Fisica Nucleare, Via Giuria 1, Torino, Italy} 
\author{J. Lange}
\affiliation{Istituto Nazionale di Fisica Nucleare, Via Giuria 1, Torino, Italy} 
\author{F. Legger}
\affiliation{Istituto Nazionale di Fisica Nucleare, Via Giuria 1, Torino, Italy}  
\author{S. Vallero}
\affiliation{Istituto Nazionale di Fisica Nucleare, Via Giuria 1, Torino, Italy}  

\date{\today}
\begin{abstract} 
We present Q-Transform Amplitude Modulation (\texttt{QTAM}), a novel, fully invertible implementation of the Constant-Q Transform (CQT) algorithm, designed to enable robust signal denoising and the disentanglement of overlapping transient events in current and, especially, next generation gravitational wave observatories. Conventional time-frequency (TF) analysis faces a fundamental dichotomy: \textit{critically sampled} transforms (e.g., standard Discrete Wavelets) are computationally efficient but lack \textit{time-shift invariance}~\cite{Kingsbury2001, Mallat2008}, limiting their efficacy for robust pattern recognition and Deep Learning applications. 
While alternative approaches such as the Dual-Tree Complex Wavelet Transform~\cite{Kingsbury2001} provide efficient approximate shift-invariance, their wavelet constructions remain tied to dyadic scale frequency tilings that are poorly matched to the simultaneous representation of compact binary chirps and instrumental glitches. Scattering transforms \cite{scatteringtransform} further improve translation stability, but do so through nonlinear modulus and averaging operations that sacrifice phase coherence and are not designed for exact signal reconstruction.
Conversely, \textit{overcomplete} transforms (e.g., standard CQT or Stationary Wavelets) provide the necessary shift-invariance and tunable frequency resolution, but their implementations generate highly redundant data volumes that are prohibitive for low-latency processing~\cite{Macas2022}. Furthermore, standard attempts to compress these dense representations rely on lossy interpolation or magnitude-only reduction, destroying the phase coherence required to reconstruct the time-domain signal. 
\texttt{QTAM} bridges this gap by employing a methodology inspired by Amplitude Modulation (AM) radio broadcasting. By modeling the Q-transform output as a slowly varying complex envelope carried by a deterministic high-frequency term, we achieve lossless data decimation via spectral shifting to baseband. 
We demonstrate that \texttt{QTAM} is linear and fully invertible, allowing exact reconstruction of the original time-series signal with machine precision while retaining the shift-invariance of dense spectrograms. Leveraging native GPU acceleration (\texttt{PyTorch}), \texttt{QTAM} achieves speedups of approximately two orders of magnitude with respect to standard implementations, enabling high-fidelity TF pipelines to operate within strict low-latency ($\mathcal{O}(1s)$) bounds. We validate the method's potential for denoising and disentanglement through tests on real gravitational wave data and simulated signal injections.\end{abstract}

\maketitle

\section{Introduction} \label{sec:intro} 
Time-frequency (TF) analysis has become a cornerstone of modern data science, offering a powerful way to understand how the frequency content of a signal changes over time. Imagine listening to a musical piece: the notes (frequencies) change as the melody progresses. Traditional methods such as the Fourier transform give you the overall frequency content of the whole piece, but they do not tell you when each note was played. TF analysis solves this problem. This approach is particularly valuable in fields where the data is complex and non-stationary allowing researchers to better understand phenomena that traditional analysis methods, such as pure time-domain or frequency-domain approaches, might miss. The earliest approaches to TF analysis, developed in the mid-20th century  \cite{Gabor1946}, involved looking at small slices of the signal at a time, like taking snapshots. This allowed for some understanding of how frequencies changed, but these methods, including the basic Short Time Fourier Transform (STFT), had limitations, such as blurring the results due to the fixed resolution \cite{QTBrown1991}. In recent years various TF representations, from the Wavelet Transform (WT) \cite{Mallat2008, Daubechies1992} to Synchrosqueezing Transform (SST) \cite{Boashash2015} and the Q-Transform (QT) \cite{QTBrown1991, Chatterji2004}, have been successfully employed to analyse data in diverse fields. Applications range from detecting transient signals in astrophysics \cite{Klimenko2016, Zevin2017} and characterizing neural oscillations in neuroscience \cite{Adeli2003}, to analyzing ECG morphology in biomedical engineering \cite{Addison2005} and modeling market volatility in finance \cite{Gencay2001}, resolving the blurring issues of standard STFT at the expense of increased computational cost.\\
Gravitational Wave (GW) and multi-messenger astronomy are prominent beneficiaries of these techniques. GWs are ripples in spacetime produced by catastrophic events, such as black hole and neutron star mergers and supernovae~\cite{Maggiore2007}. These signals often exhibit a characteristic ‘‘chirp'' (frequency increase) that is readily discernible in spectrograms, providing an evolutionary fingerprint of astrophysical phenomena. TF analysis is critical to detect signals from unmodeled sources (bursts)~\cite{cWB, burst, Bayesinf, Chatterji2004} and for distinguishing astrophysical signals from environmental noise and instrumental glitches~\cite{Virgotools, Omicron, GSpyNetTree, GravitySpy, t-SNE, Gengli}. 
The QT has emerged as the standard diagnostic tool in this field. By adapting its TF resolution to match the geometric progression of frequency, the QT is uniquely suited to capture the morphology of compact binary coalescences, producing 2D spectrograms that visually highlight transient phenomena with high fidelity~\cite{Chatterji2004, GWpy2021}. Such spectrograms enable the application of Deep Neural Networks (DNNs), such as Convolutional Neural Networks (CNNs), for tasks ranging from glitch classification~\cite{GSpyNetTree, GravitySpy, t-SNE, Gengli} to parameter estimation via Normalizing Flows~\cite{GramaxoFreitas2024}. However, a distinct dichotomy exists within the LIGO-Virgo-KAGRA (LVK) data analysis landscape. While unmodeled searches operating in low-latency ($\mathcal{O}(1s)$) (a constraint strictly dictated by the need to trigger multi-messenger alerts and guide electromagnetic telescopes to transient events before they fade) successfully employ sparse, lower-resolution wavelet decompositions (e.g. coherent WaveBurst (cWB) \cite{cWB}) to identify signals from energetic pixels, the high-resolution, dense spectrograms essential for detailed ML morphology recognition are largely relegated to offline visualization or classification.
This dichotomy exists because generating and processing high-resolution 2D data is historically computationally prohibitive for the low-latency bounds required for alerts~\cite{Macas2022, lowlatency}. Furthermore, standard techniques for data compression (in order to reduce spectrograms to manageable sizes) such as interpolation or downsampling are inherently lossy. Such processes destroy phase coherence, introduce artifacts (e.g., negative energy values in the case of interpolation), and render the transformation non-invertible, thereby preventing the integration of any processed 2D data back into precision time-domain pipelines.\\
These limitations present a critical bottleneck for third-generation observatories, such as the Einstein Telescope (ET) and Cosmic Explorer (CE)~\cite{ET,CE}, as such observatories are forecast to witness a dramatic increase in the number of detected GW events, from roughly 100 to as many as 100,000 per year \cite{ET, CE}. Furthermore, the observed signals will have a significantly longer duration (up to hours or days for low-mass binaries) \cite{ET,CE}. This unprecedented data volume and signal duration will lead to a much higher likelihood of overlapping signals \cite{ET,CE}, making TF representation of the data crucial for the GW community: by leveraging the sparsity of transient signals in the TF plane (where distinct events follow separate evolutionary track), this approach enables the precise localization and separation of intertwined signals, a task that traditional 1D methods struggle to achieve in an unbiased way \cite{ETDataAnalysis2025, overlap}.
To fully translate this analytical advantage into a viable pipeline and strictly avoid computational bottlenecks, future algorithms require key attributes: they must efficiently exploit cutting-edge hardware \cite{Bagnasco2024} (such as GPUs, shared memory CPU/GPU architectures and HPC centers), operate within low-latency bounds, and produce reasonable sized outputs without compromising data faithfulness (i.e., not relying on standard lossy downsampling methods that discard phase, nor not time-shift invariant critical sampling). Crucially, the representation must be fully invertible \cite{Qptransform} to allow for robust hybrid time and TF analyses strategies, and must bring the data into a format compatible with ML pipelines (CNNs and Transformers \cite{cnn, transformers}) that are likely to play an ever-growing role in detection and parameter estimation in the GW data analysis of the near future \cite{Cuoco_2021}. We address all the above issues with a novel implementation of the QT algorithm that we name \texttt{QTAM} (Q-Transform Amplitude Modulation). \texttt{QTAM} resolves the fundamental trade-off between representational fidelity, data volume, and computational latency. By mathematically demodulating the spectral coefficients to baseband, \texttt{QTAM} retains the shift-invariance and tunable resolution of dense, overcomplete transforms (essential for robust feature extraction) while reducing data redundancy to the Nyquist-Shannon sampling limit defined by the signal bandwidth. 
Rather than introducing a new transform in a mathematical sense, \texttt{QTAM} combines the established theory of constant-Q tilings with a baseband decimation framework, resulting in an optimal representation that directly translates to speed and enables high-fidelity analysis within the strict latency bounds required for global alerts.\\
We first outline the mathematical formulation of the QT and its current implementations adopted by the LVK community (Sec.~\ref{sec:sota}). We then introduce the \texttt{QTAM} algorithm (Sec.~\ref{sec:qtam}), detailing the amplitude demodulation technique that enables lossless downsampling. In Sec.~\ref{sec:results}, we compare the computational performance of \texttt{QTAM} against state-of-the-art implementations. Finally, in Sec.~\ref{sec:denoise}, we demonstrate the efficacy of \texttt{QTAM} for GW denoising and signal disentanglement on real and simulated data.
\section{State of the art}
\label{sec:sota} 
\begin{figure}[h!] 
    \centering
    \includegraphics[width=0.7\linewidth]{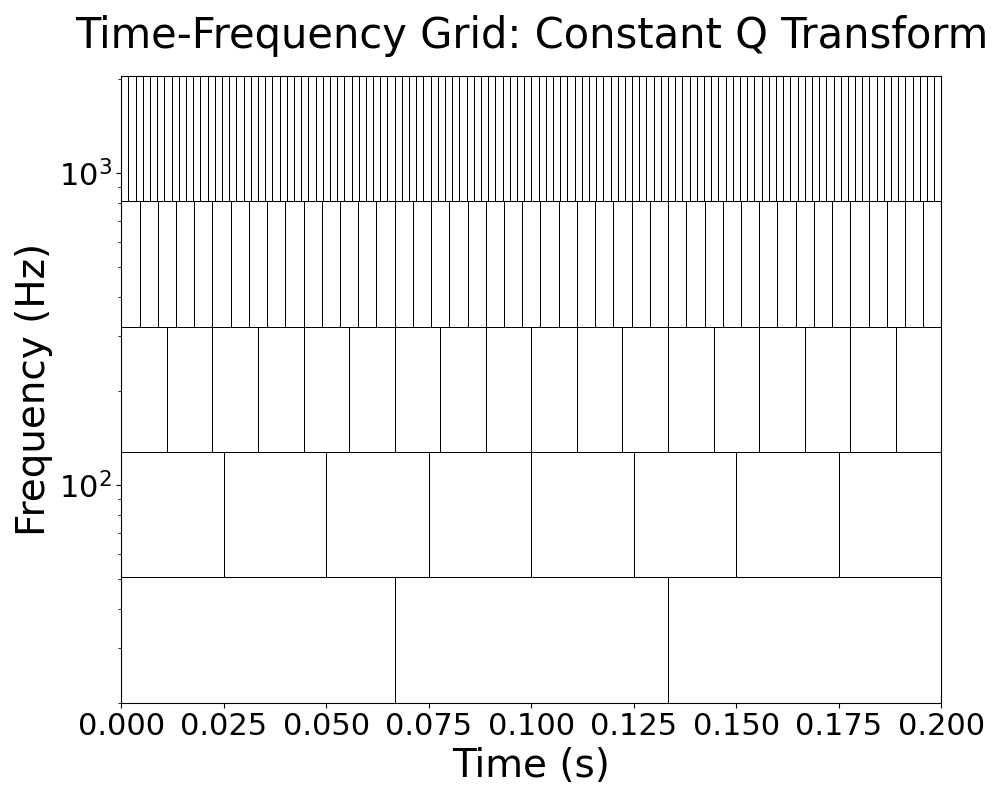}
    \caption{The TF plane on which the CQT is computed. Frequencies range from 20 to 2048 Hz over a 0.2s interval. The horizontal frequency bands are referred to as \textit{tiles}, exhibiting varying temporal resolutions.}
    \label{fig:TF_Grid}
\end{figure}

The Constant-Q Transform (CQT) maps a time-series signal into a spectral representation where the frequency resolution is geometrically related to the frequency~\cite{QTBrown1991}. This implies a non-uniform tiling of the TF plane: low-frequency bins possess high spectral resolution but poor temporal precision, while high-frequency bins sacrifice spectral detail for sharp temporal localization. Formally, it is constructed by convolving the signal with a bank of windowed complex exponentials, where the duration of each window is inversely proportional to its central frequency $f_k$. \\
For a set of $n$ frequency bands per octave, the transform maintains a constant quality factor $Q$, defined as the ratio of the central frequency to the bandwidth: $Q=f_{k}/\Delta f_{k}$. Consequently, the bandwidths follow a geometric progression:
\begin{equation}
    \Delta f_{k}= 2^{1/n} \Delta f_{k-1}= 2^{k/n} \Delta f_{min}
\end{equation}
Governed by the uncertainty principle $\Delta f \Delta t \geq 1/4\pi$, this variable frequency scaling necessitates a TF grid with non-uniform time resolution, as illustrated in Fig.~\ref{fig:TF_Grid}.

The continuous CQT of a signal $x(t)$ for a frequency $f_k$ is computed as:
\begin{equation} \label{eq:CQT}
    T (\tau, f_k, Q)= \int_{-\infty}^{+\infty} x(t) w(t-\tau, f_k, Q) e^{-2\pi i f_k t} dt
\end{equation}
where $w(t, f_k, Q)$ is a window function whose length scales with $Q/f_k$. The discrete implementation for a frequency bin $k$ is given by:
\begin{equation} \label{eq:Discrete_CQT}
    T [m, k, Q]= \frac{1}{N_k}\sum_{n=0}^{N_k-1} x[m+n] w[n, k] e^{-2 i\pi  n \frac{Q}{N_k}}
\end{equation}
where $N_k$ represents the window length in samples for bin $k$. The resulting output is a complex-valued 2D matrix. The row vector corresponding to a specific frequency $f_k$ is referred to as a \textit{tile}, possessing a temporal duration and resolution distinct from its neighbors.\\
While the continuous Q-transform is not, in general, invertible under standard wavelet theory because its kernels do not satisfy the admissibility condition, discrete implementations may still admit reconstruction in principle when the window bank forms a complete frame with nonzero spectral coverage (see Appendix \ref{app:invertibility}).\\
\\
The standard \texttt{GWpy} \cite{GWpy2021} implementation of the CQT, widely used for visualization of GW data, computes the transform tiles $T(\tau,f,Q)$ by leveraging the Convolution Theorem and operating primarily in the frequency domain: 
\begin{equation} \label{eq:gwpy_convolution_prd} \int x(\tau) w(t - \tau,f, Q) d\tau = \mathcal{F}^{-1} \left[ \tilde{X}(f) \times \tilde{W}(f,Q) \right]
\end{equation} 
where $\tilde{x}(f)=\mathcal{F}[x(t)]$ is the Fourier transform of the signal  and   $\tilde{W}(k,f,Q)=\mathcal{F}[w(t,f,Q)]$ that of the window. The implementation is discrete, and uses a basis of bisquare windows, directly defined in frequency domain for a computational speedup, which are normalized to ensure energy preservation across the resultant tiles:
\begin{equation} \label{eq:gwpy_window_and_normalization} 
\tilde{W}[k,f] = C_{f} \cdot \tilde{W'}[k,f], \quad \quad \quad C_{f} = \frac{1}{\sum_{k} (\tilde{W'}[k,f])^2},\end{equation}
where:
\begin{equation}
\label{eq:gwpy_window_formula}
\tilde{W}'[k, f] = \left(1 - \left(\frac{k - k_c}{\Delta k_f}\right)^2\right)^2, \quad \text{for } k \in \mathcal{K}_{k_c}
\end{equation}
and the frequency-dependent support 
$\mathcal{K}_{k_c}$ and bandwidth $\Delta k_{f}$ are defined as:
\begin{equation} \label{eq:support}
\mathcal{K}_{k_c} = \Bigl\{ k \in \mathbb{Z} \ \Big| \ |k - k_c| \le \Delta k_f \Bigr\}, \quad \Delta k_f = \lfloor N_f / 2 \rfloor
\end{equation}
with $k_{c}=fN_t$ being the central frequency index and $N_t$ the duration of the signal. 
Because this basis is defined over a frequency-dependent support, the resulting Q-tiles possess varying temporal lengths. To produce the uniform 2D spectrograms required for visualization, the \texttt{GWpy} implementation subjects the data to three distinct, destructive operations, each of which independently renders the discrete transformation implementation strictly non-invertible. First, it returns only the squared magnitude $|T(\tau, f, Q)|^2$, completely discarding the complex phase information necessary for signal reconstruction. Second, it applies median normalization (effectively whitening the tiles), thereby destroying the absolute amplitude scaling of the signal. Finally, it employs lossy interpolation (via bicubic splines) to map the variable-length tiles onto a fixed grid, corrupting the native TF support. These irreversible steps prevent the integration of 2D representations back into phase-sensitive time-domain analyses. Computationally, this implementation is currently restricted to CPU execution, largely relying on sequential operations.

The \texttt{ml4gw} CQT implementation \cite{ml4gw} represents a shift toward high-efficiency computation tailored for ML applications. Built upon the \texttt{PyTorch} framework, it leverages native GPU acceleration, parallelization and vectorization to achieve speedups of $\mathcal{O}(10^2)$ to $\mathcal{O}(10^3)$ over CPU-based implementations. As the core algorithm mathematically mirrors \texttt{GWpy} (yielding identical numerical results up to machine precision) it inherits the standard implementation's limitation of discarding phase to produce magnitude-only spectrograms for CNN inputs. However, \texttt{ml4gw} offers greater flexibility by providing access to the raw, non-interpolated tiles and allowing users to select between interpolation methods (bi-spline vs. bicubic) to optimize the trade-off between output smoothness and computational latency.

Both the theoretical inversion limitations of the classical Q-transform and the practical irreversibility of the software implementations of \texttt{GWpy} and \texttt{ml4gw} are rigorously addressed by the Wavelet Q-transform (WQT) implementation~\cite{Qptransform}. Unlike standard discrete implementations that discard phase and rely on lossy interpolation, this framework retains the full complex information by treating the transform as a Continuous Wavelet Transform (CWT) defined via a family of energy-normalized complex wavelets:
\begin{equation} \label{eq:qtransform_wavelet} 
    w^*(t; \tau, \nu, Q) = \left( \frac{8\pi \nu^2}{Q^2} \right)^{1/4} e^{ - \left( \frac{2\pi \nu(t-\tau)}{Q} \right)^2 } e^{-2\pi i \nu (t-\tau)}
\end{equation} 
whose Fourier transform has constant support that covers the full spectrum of the signal, thus eliminating the need for interpolation in order to achieve a uniform 2D spectrogram. Despite violating the \textit{admissibility condition} (vanishing integral) required for standard CWT inversion \cite{Daubechies1992, waveletinv}, such a basis allows for a specialized inversion formula to recover the signal $x(t)$ not by simple projection, but by integrating the time-derivative of the transform coefficients $T(t, \nu, Q)$:
\begin{equation} \label{eq:qtransform_inversion} 
x(t) = \frac{2} {\mathcal{C}_Q} \cdot \mathrm{Re} \left[ \int_{0}^{+\infty} \frac{d\nu}{i \pi \nu \sqrt{2\pi\nu^2}} \frac{\partial}{\partial t} T (t, \nu, Q) \right]
\end{equation} 
where $\mathcal{C}_Q = (2/\pi Q^2)^{1/4} \mathrm{Re} [ \mathrm{erf} ( Q / 2 ) ]$. The authors also present an extension in~\cite{Qptransform}, the Wavelet Qp-Transform (WQpT), which introduces a chirp parameter $p$ into the wavelet definition to maximize sparsity for compact binary coalescences, though the inversion logic remains analogous.\\
Despite its theoretical rigor, the WQT implementation \cite{virtuoso_zenodo,UnoaccasoRepo} faces significant performance bottlenecks. The current official release of the code \cite{virtuoso_zenodo} runs on CPU only and lacks batch vectorization, necessitating sequential data processing. Even recent experimental GPU-based efforts \cite{UnoaccasoRepo} remain unoptimized and similarly non-vectorized, relying on multiprocessing to manage computational loads. Moreover, the strategy employed to circumvent tile interpolation introduces a computational overhead. To ensure a visually satisfactory representation without interpolation, the method adopts a continuous approach where wavelet windows are defined over the entire Fourier support. This prevents the reduction of temporal resolution at lower frequencies, necessitating a highly redundant tiling density to preserve visual continuity\cite{Qptransform}. This results in output matrices of often impractical size.

Omicron \cite{Omicron} is a highly optimized CQT implementation that is fast enough to be fully integrated into the LVK online data analysis pipelines for the low-latency detection and characterization of GW transients and instrumental glitches. The algorithm computes a stack of CQT planes across a range of Q values. The number of planes is spaced logarithmically to ensure that fractional energy loss (mismatch) between adjacent planes remains below a user-defined threshold.
Similar to the \texttt{GWpy} implementation, Omicron uses a bisquare window basis implemented directly in frequency domain. To synthesize a single TF map from these overlapping planes, Omicron avoids the smoothing effects of interpolation. Instead, it projects the multi-Q data onto a pixel grid: for each pixel, the algorithm identifies all overlapping tiles across the full range of Q values and selects the single tile with the highest amplitude squared. This maximization strategy automatically selects the optimal Q-value for a given signal feature, maximizing sensitivity to transients while ensuring the output represents discrete physical data rather than interpolation artifacts. Computationally, Omicron is entirely CPU-based, relying on optimized libraries without leveraging GPU acceleration.\\
Coherent WaveBurst (\texttt{cWB})~\cite{cWB} represents the current gold standard for low-latency unmodeled burst detection. It employs the Wilson–Daubechies–Meyer wavelet transform (WDM) wavelet transform~\cite{Daubechies1992}, which is orthogonal and critically sampled. While this ensures high computational efficiency and good sensitivity to chirping signals, the resulting TF representation is not shift-invariant: small shifts in the input signal can alter the wavelet coefficients. To address this limitation, an overcomplete extension, \texttt{cWB-XP}~\cite{cWB-XP}, has been developed, which stacks multiple continuous, overlapping TF layers to improve robustness to small time shifts. In contrast, overcomplete transforms such as the Constant-Q transform naturally provide shift-invariance through redundancy, making them particularly suitable for deep learning architectures. Furthermore, the rigid WDM tiling in standard \texttt{cWB} limits flexibility in matching diverse signal morphologies, whereas the CQT allows continuous tuning of the Q-factor to adapt not only to chirping gravitational-wave signals but also to transient noise features and other nonstationary structures.

This survey highlights a fundamental dichotomy in the current landscape. Overcomplete methods offer the native shift-invariance and phase coherence essential for robust feature extraction, yet their prohibitive data redundancy renders them computationally intractable for high-dimensional 2D neural networks. Existing attempts to compress these representations sacrifice fidelity through lossy interpolation, breaking the invertibility required for physical interpretation. Conversely, critically sampled methods achieve efficiency but lack intrinsic shift-invariance. While pipelines like \texttt{cWB} recover robustness by aggregating multiple resolution layers, this renders invariance an emergent property of the algorithmic search rather than an intrinsic property of the data representation. Such a representation is ill-suited for standard Deep Learning architectures, as it requires models to learn complex cross-layer correlations to recover stability, rather than exploiting the native geometric coherence provided by overcomplete transforms. 

More general invertible TF representations can, in principle, be constructed within the framework of frame theory, including Gabor and nonstationary Gabor transforms \cite{grochenig2001foundations,velasco2011constructing, holighaus2013framework}, which provide flexible tilings and exact reconstruction. However, these approaches are not currently adopted in data analysis pipelines within the \texttt{LVK} collaboration, largely due to their computational cost and the lack of GPU-optimized implementations compatible with low-latency constraints. In parallel, demodulation techniques such as heterodyning \cite{oppenheim1999signals} are widely used in signal processing to shift band-limited signals to baseband, enabling sampling at rates determined by the local bandwidth rather than the full signal bandwidth. In GW data analysis, similar strategies are routinely employed in continuous-wave searches to isolate narrowband signals and reduce computational cost \cite{cornish2010fast,dupuis2005bayesian}. However, these ideas have not been applied to overcomplete TF representations such as the Q-transform, where coefficients are typically computed and stored at the full sampling rate.

The field therefore lacks a unified framework capable of compactly representing overcomplete transforms: retaining the high-resolution geometry of the Q-transform while reducing its data footprint to the minimum sampling rate dictated by the signal bandwidth, rather than the global sampling rate.

In the next section, we introduce \texttt{QTAM}, a novel algorithm built on the Q-transform framework already used in the gravitational-wave field, which resolves this tension by demodulating each TF component to baseband. This enables a minimal-sampling representation of the shift-invariant transform while preserving exact reconstruction and maintaining compatibility with low-latency, GPU-accelerated pipelines.

\section{QTAM: Amplitude modulation.}\label{sec:qtam}

In this section we present \texttt{QTAM}, a novel discrete QT algorithm designed to address the conflicting requirements described in Sec. \ref{sec:sota} simultaneously. By ensuring linearity, full invertibility, and GPU-accelerated performance, \texttt{QTAM} enables the seamless integration of 2D denoising and analysis back into time-domain pipelines within strict low-latency bounds ($\mathcal{O}(1)\text{s}$). Our approach begins by tackling the most pressing challenge: the compression of the complex signal. Drawing inspiration from Amplitude Modulation (AM) radio broadcasting, we reinterpret the Q-transform output not as a static image, but as a dynamic transmission problem, separating the slowly varying signal envelope from its carrier frequency to achieve mathematically lossless decimation. 

The conceptual foundation of \texttt{QTAM} mirrors the solution to a classic engineering challenge in early telecommunications: the transmission of low-frequency audio signals. In radio broadcasting, the direct transmission of audio frequencies (20 Hz - 20 kHz) is physically impractical because efficient antenna radiation requires the antenna length to be proportional to the signal's wavelength ($\lambda = c/f$). For a 1 kHz audio tone, the required antenna would need to be hundreds of kilometers long. 
Historically, this problem was solved through the application of Amplitude Modulation (AM), a concept rooted in the wired "undulatory current" of Alexander Graham Bell \cite{Bell1876} and the wireless experiments of Roberto Landell de Moura \cite{Landell1901}, before being adapted for continuous-wave radio by Reginald Fessenden \cite{Fessenden1902, HaykinCommunicationSystems} and later perfected for high-efficiency reception and demodulation by Edwin Armstrong \cite{Armstrong1914}. Fessenden realized that instead of transmitting the low-frequency signal directly, one could use it to modulate the amplitude (envelope) of a much higher frequency "carrier" wave. This shifted the spectral information from baseband to the high-frequency domain, where reasonable-sized antennas could operate efficiently. If the receiver is tuned to this specific carrier frequency, it can reverse the process—demodulating the incoming wave to perfectly recover the original low-frequency audio.

\begin{figure}[h!] 
    \centering
    \includegraphics[width=0.9\linewidth]{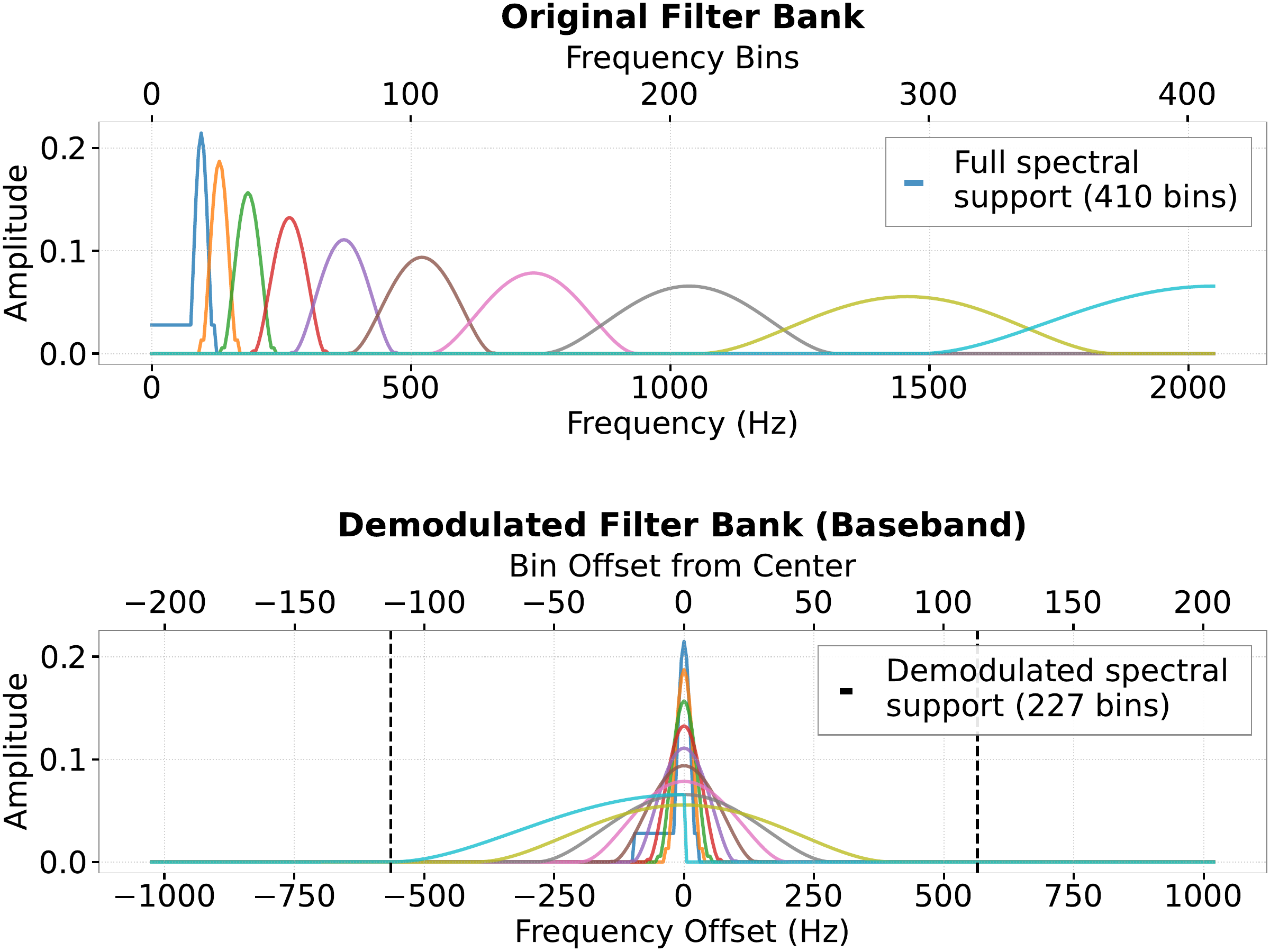}
    \caption{Schematic of spectral demodulation in \texttt{QTAM}. \textit{Top}: The standard CQT filter bank where windows are centered at their absolute frequencies $f_k$. The large spectral support necessitates a high sampling rate to resolve the carrier oscillations. \textit{Bottom:} The same windows after being circularly shifted to baseband ($0 Hz$). By removing the carrier frequency offset, the spectral support is compressed to the signal's inherent bandwidth, allowing for lossless downsampling.}
    \label{fig:demod}
\end{figure}

\texttt{QTAM} addresses the inverse problem: while radio broadcasting modulates low-frequency information onto a high-frequency carrier to enable transmission, standard CQT analysis inadvertently retains this ‘‘carrier'' structure, transforming it into a computational bottleneck. Mathematically, a CQT tile centered at frequency $f_k$ can be decomposed into a slowly varying complex envelope $Y_k(t)$ modulated by a deterministic, rapidly oscillating carrier $e^{j2\pi f_k t}$:
\begin{subequations}
\begin{align}
T_k(t) &= Y_k(t) \cdot e^{j2\pi f_k t} \\
Y_k(t) &= A_k(t) \cdot e^{j\phi_k(t)}
\end{align}
\end{subequations}
where $A_k(t)$ represents the physical amplitude and $\phi_k(t)$ the relative phase evolution. The inefficiency arises because preserving the absolute oscillation of the carrier requires a sampling density in the output proportional to $f_k$ to satisfy the Nyquist criterion, even though the information content within $Y_k(t)$ varies much more slowly. \texttt{QTAM} resolves this by recognizing that the physical signal is defined by the relative spectral variations within the window, not its absolute frequency position. By circularly shifting the spectrum to baseband, the algorithm effectively demodulates the signal, removing the carrier oscillation as show in Fig. \ref{fig:demod}.
\\
This reduces the maximum frequency to the half-bandwidth of the largest window, enabling aggressive, lossless decimation that scales with the information density rather than the central frequency. As the effective Inverse Fast Fourier Transform (IFFT) size is changed, a renormalization factor of $\sqrt{N_{out}/N_{in}}$ is applied to satisfy Parseval's theorem, ensuring that the physical envelope amplitude $A_k(t)$ remains invariant to the change in sampling density. As the baseband shifted window basis is symmetrically zero padded to match the bandwith of the largest window (without altering the frequency content of the signal), the resulting tiles all share the same length, no further interpolation nor downsampling is required, and no artefacts are introduced. Since the carrier frequency 
$f_k$ is deterministic, this transformation is fully reversible. The absolute phase evolution can be recovered by re-modulating the downsampled envelope $Y_k(t)$ with the carrier $e^{j2\pi f_k}$. This yields a faithful complex representation of the original CQT tile while preserving the compact array size. However, as shown in the next section, avoiding this re-modulation and processing directly in baseband holds the potential to yield superior results in ML applications.\\
\texttt{QTAM} is a \textit{linear} operator. Formally, the transformation 
$\mathcal{T}:x(t)\rightarrow Y(t,f)$ satisfies 
$\mathcal{T}(\alpha x_1 + \beta x_2) = \alpha \mathcal{T}(x_1) + \beta \mathcal{T}(x_1)$. This property follows directly from the linearity of the underlying convolution and demodulation operations. Linearity is critical for downstream ML tasks, particularly denoising such as in \cite{GlitchFlow2024}, as it ensures that additive noise in the time domain remains additive in the TF space, preventing the introduction of cross-term artifacts common in quadratic (spectrogram) representations.

The invertibility of \texttt{QTAM} is grounded in \textit{Discrete Frame Theory} \cite{Mallat2008}, which naturally aligns with GW data, which is discrete itself. Unlike the CWT formalism, which requires the basis functions to satisfy the \textit{admissibility condition}, this discrete approach relies strictly on the completeness of the window bank $\Psi=\{ W_{k} \}$: provided that the spectral coverage satisfies the frame bounds, specifically that the aggregate power spectrum is non-zero and continuous across the frequency range, the signal $x[n]$ can be perfectly reconstructed from the coefficients $Y_k[n]$ via the canonical dual frame (see Appendix \ref{app:invertibility}). In practice, this inversion is computed efficiently in the frequency domain:
\begin{equation}\label{eq:qtam_canonical_inverse}
x[n] = \mathcal{F}^{-1} \left\{ \frac{\sum_{k} \tilde{T}_k[f] \cdot \tilde{W}_k^{*}[f]}{\sum_{k} |\tilde{W}_k[f]|^2} \right\}
\end{equation}
where $\tilde{T}_k$ represents the spectrum of the $k$-th tile after being shifted back from baseband to its original position $f_k$ (applying the necessary amplitude correction $\sqrt{N_{in}/N_{out}}$), and the denominator $\sum_k \vert \tilde{W}_k[f]\vert^2$ acts as a pre-computed normalization vector (the frame operator) that corrects for the overlapping energy of the non-orthogonal windows.\\
Additionally, in applications where the amplitude demodulation step is bypassed, \texttt{QTAM} offers an alternative inversion method exploiting the maximum redundancy of the basis. Users may optionally initialize the frequency-domain window basis with a non-zero regularization term $\epsilon$ in the stopband region:
\begin{equation}
\label{eq:qtam_window_formula}
\tilde{W}'[k,f] =
\begin{cases}
\left(1 - \left(\frac{k - fN_{t}}{\lfloor N_{f}/2 \rfloor}\right)^2\right)^2 & \text{if } \left|k - fN_{t}\right| \le \lfloor N_{f}/2 \rfloor \\
\epsilon & \text{otherwise,} \quad 0<\epsilon\ll 1
\end{cases}
\end{equation}
By selecting a sufficiently small $\epsilon$~\footnote{i.e. of the order of machine precision. Note that too small values might cause under/overflow or approxiamation errors, effectively altering the values of the Qtransform}  this modification has negligible impact on the forward CQT computation, yet renders the window function $\tilde{W}[k]$ strictly non-zero and element-wise invertible. Consequently, the full time-domain signal can be algebraically recovered from any single tile via direct spectral division:
\begin{equation}
x[n] = \mathcal{F}^{-1} \left\{ \frac{\tilde{T}_k[f]}{\tilde{W}_k[f]} \right\}\end{equation}

Built natively in \texttt{PyTorch}, \texttt{QTAM} achieves remarkable speed performances (see Sec.~\ref{sec:results}) by implementing demodulation and decimation as vectorized tensor operations suitable for high-throughput GPU execution. Beyond raw performance, the framework prioritizes flexibility, allowing users to customize the analysis basis with various window functions (e.g., \textit{Bisquare, Tukey, Planck-taper}) and define arbitrary frequency grids, ranging from standard logarithmic spacing to specific user-defined bins. Crucially, the output temporal resolution is fully adjustable; the algorithm automatically calculates the minimum sampling rate required to avoid aliasing, ensuring lossless representation. Additionally, \texttt{QTAM} supports a configurable maximum window clamp, effectively transitioning from CQT to STFT behavior at high frequencies (typically $\mathcal{O}(10^3Hz)$), which enables even more aggressive downsampling limits (see Appendix~\ref{app:implementation} for full implementation details). 

In the following section we benchmark \texttt{QTAM}'s computational throughput and illustrate its application to real GW data, with a specific focus on how spectral demodulation impacts phase coherence and enables high-fidelity reconstruction even under aggressive downsampling.
 
\section{Results}\label{sec:results}
We evaluated \texttt{QTAM}'s fidelity and computational efficiency using data from the Advanced LIGO and Advanced Virgo detectors. This evaluation focuses first on the exactness of the demodulation and inversion techniques using the specific case of GW150914, followed by a large-scale benchmarking of processing speed against state of the art implementations.

\begin{figure}[h!]
\centering
\includegraphics[width=0.99\linewidth]{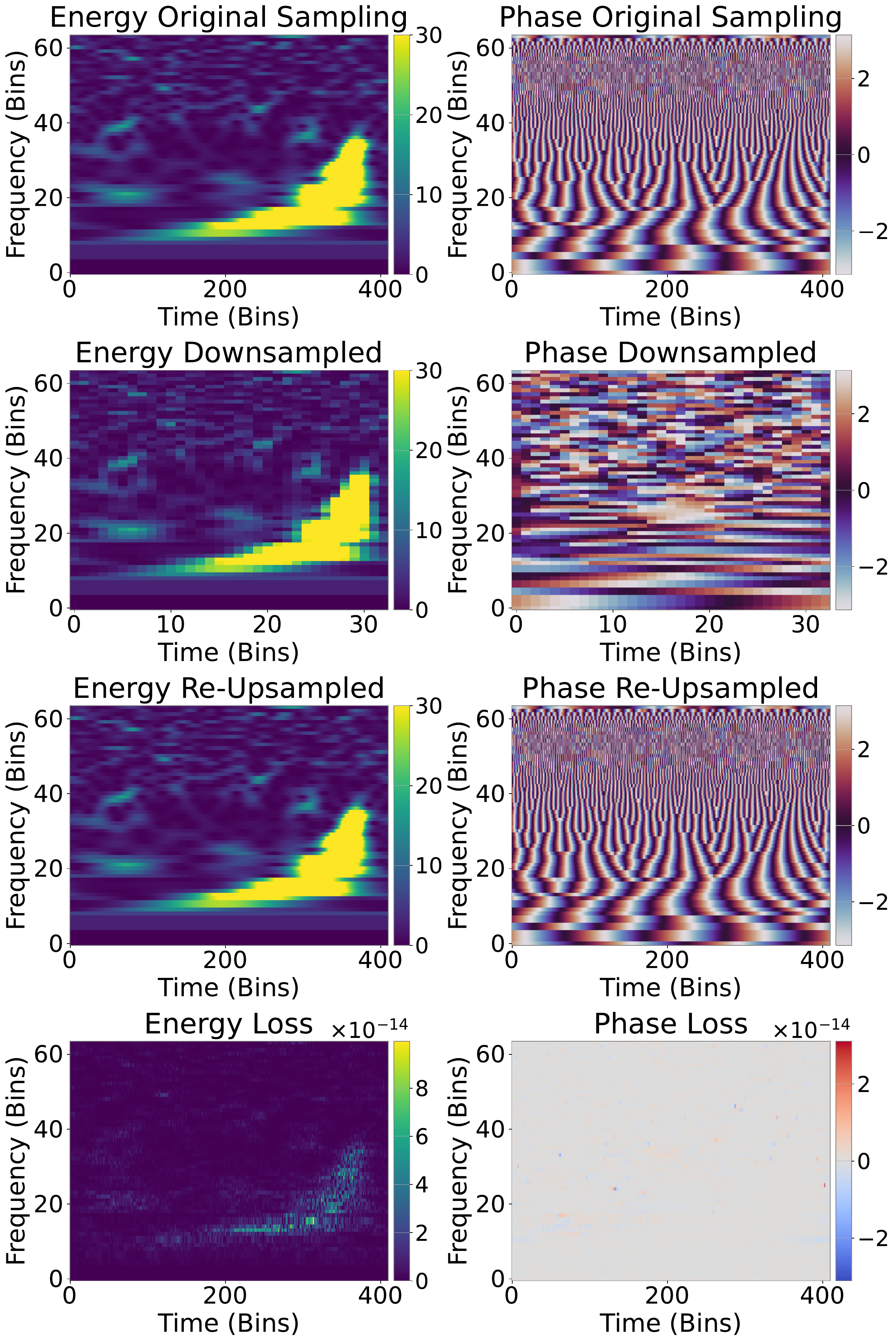}
\caption{Demonstration of information preservation under volume compression. Panels (a) and (b) show the standard resolution
$(64 \times 410)$. Panels (c) and (d) show the compressed representation ($64 \times 33$). Panels (e) and (f) show the re-upsampled reconstruction. The residuals between the original and reconstructed matrices are negligible $\mathcal{O}(10^{-14})$, confirming that the compression discards only redundant data.}
\label{fig:downsampled}
\end{figure}

\begin{figure}[h!]
\centering
\includegraphics[width=0.9\linewidth]{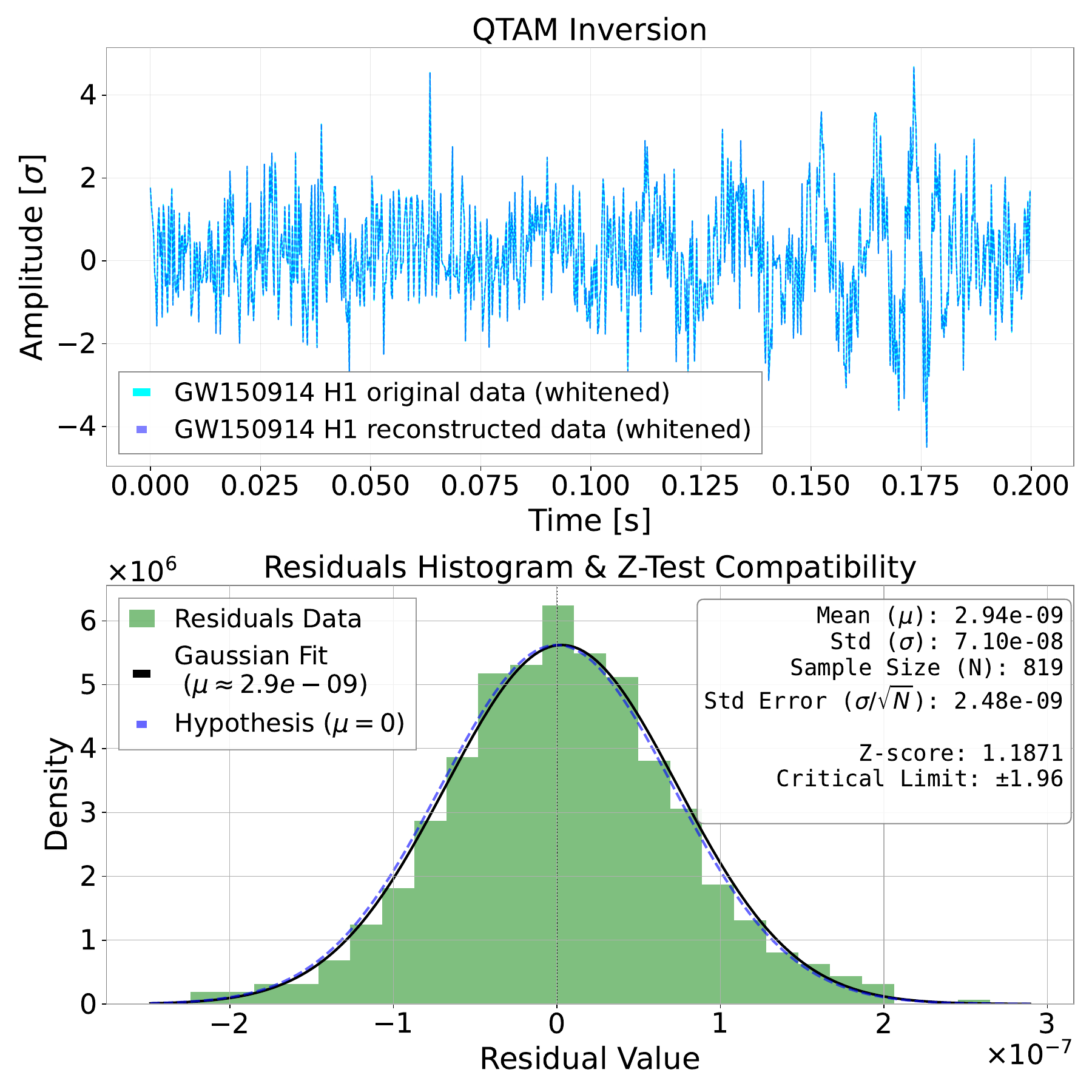}
\caption{\textit{(Top)} Overlay of the original GW150914 (H1) whitened strain data (cyan) and the signal reconstructed via the inverse \texttt{QTAM} transformation (blue). The signals are visually indistinguishable. \textit{(Bottom)} Histogram of the reconstruction residuals ($x_{\mathrm{inv}} - x_{\mathrm{orig}}$) overlaid with a Gaussian fit (black solid line) and the theoretical zero-mean hypothesis (blue dashed line). The residuals follow a Gaussian distribution with a standard deviation of $\sigma \approx 7.10 \times 10^{-8}$. A Z-test was performed to check for zero-mean compatibility ($Z \approx 1.19$), confirming that the reconstruction is accurate up to numerical precision.}
\label{fig:inversion}
\end{figure}

\begin{figure}[h!] 
    \centering
    \includegraphics[width=0.99\linewidth]{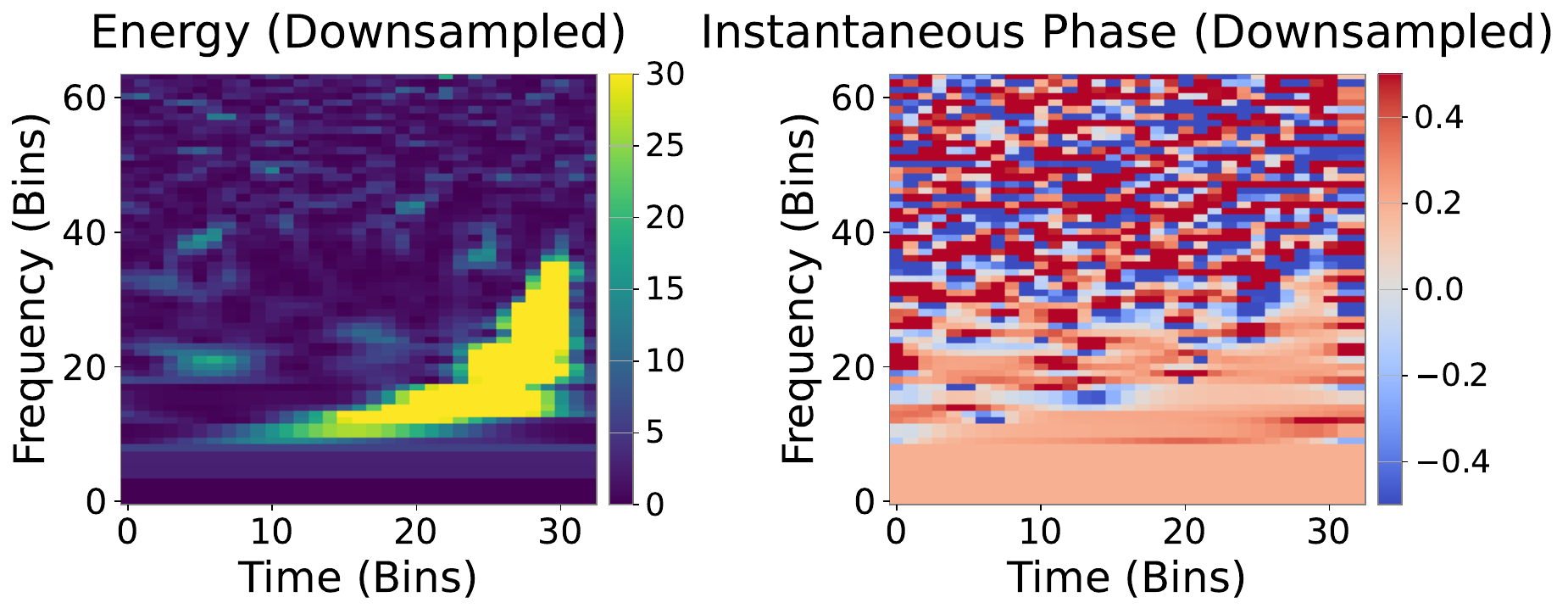}
    \caption{Demodulated signal features of (H1) GW150914. \textit{Left}: Energy spectrogram in the compressed baseband representation ($64 \times 33$ bins). \textit{Right}: Instantaneous frequency, computed as the time-derivative of the unwrapped phase. By removing carrier oscillations, the demodulated phase derivative explicitly reveals the signal's "chirp" morphology as a coherent track.}
    \label{fig:inverted}
\end{figure}

To validate the accuracy of \texttt{QTAM}, we analyzed a 0.2s segment of LIGO Hanford detector data centered on the GPS time of the event GW150914 \cite{Trovato:2019pr}. The initial computation of the full \texttt{QTAM} yields amplitude and phase matrices of size 
$64\times 410$ (frequency $\times$ time bins). On an NVIDIA H100 GPU, this transformation is completed in approximately 0.02s. A primary advantage of the demodulation technique is its ability to achieve significantly more  compact representation while preserving physical signal integrity. By applying amplitude demodulation, we downsampled the TF representations from $64\times 410$ to $64 \times 33$, which corresponds to a compression factor of approximately 12, reducing the data volume to near the Nyquist-Shannon sampling limit imposed by the 145 Hz bandwidth of the transform's largest window $n=\lceil N_t\cdot\Delta f_{max} \rceil +1$.
To confirm that this volume compression is lossless regarding physical information, we reversed the process by up-sampling the compressed images to their original dimensions. As illustrated in Fig. \ref{fig:downsampled}, the residuals between the original and reconstructed images are $\mathcal{O}(10^{-14})$ . These residuals are attributable solely to (squared) single-precision numerical rounding, confirming that the downsampled state retains all physical information required to perfectly reconstruct the signal.
We further validated the method's invertibility by applying the \texttt{QTAM} inversion algorithm to the full amplitude and phase matrices. By comparing the reconstructed time series to the original whitened input data (Fig. \ref{fig:inversion}), we observe residuals comparable to machine error size $\mathcal{O}(10^{-7})$. This precise invertibility ensures that \texttt{QTAM} allows for seamless transitions between the compressed TF domain and the pure time domain.\\
Beyond compression, phase demodulation provides a mechanism to distill physical features from the carrier signal. In standard representations, the physical evolution of the wave's phase is often obscured by carrier frequency oscillations. By factoring out these frequencies, the underlying signal structure becomes explicit. Fig. \ref{fig:inverted} demonstrates that the wave's morphology can be inferred directly from the instantaneous phase (the time derivative of the phase).
This suggests that working with demodulated, compressed data may be advantageous for machine learning applications, as NNs can more readily extract physical features when the carrier is absent.

\begin{figure}[h!] 
    \centering
    \includegraphics[width=0.99\linewidth]{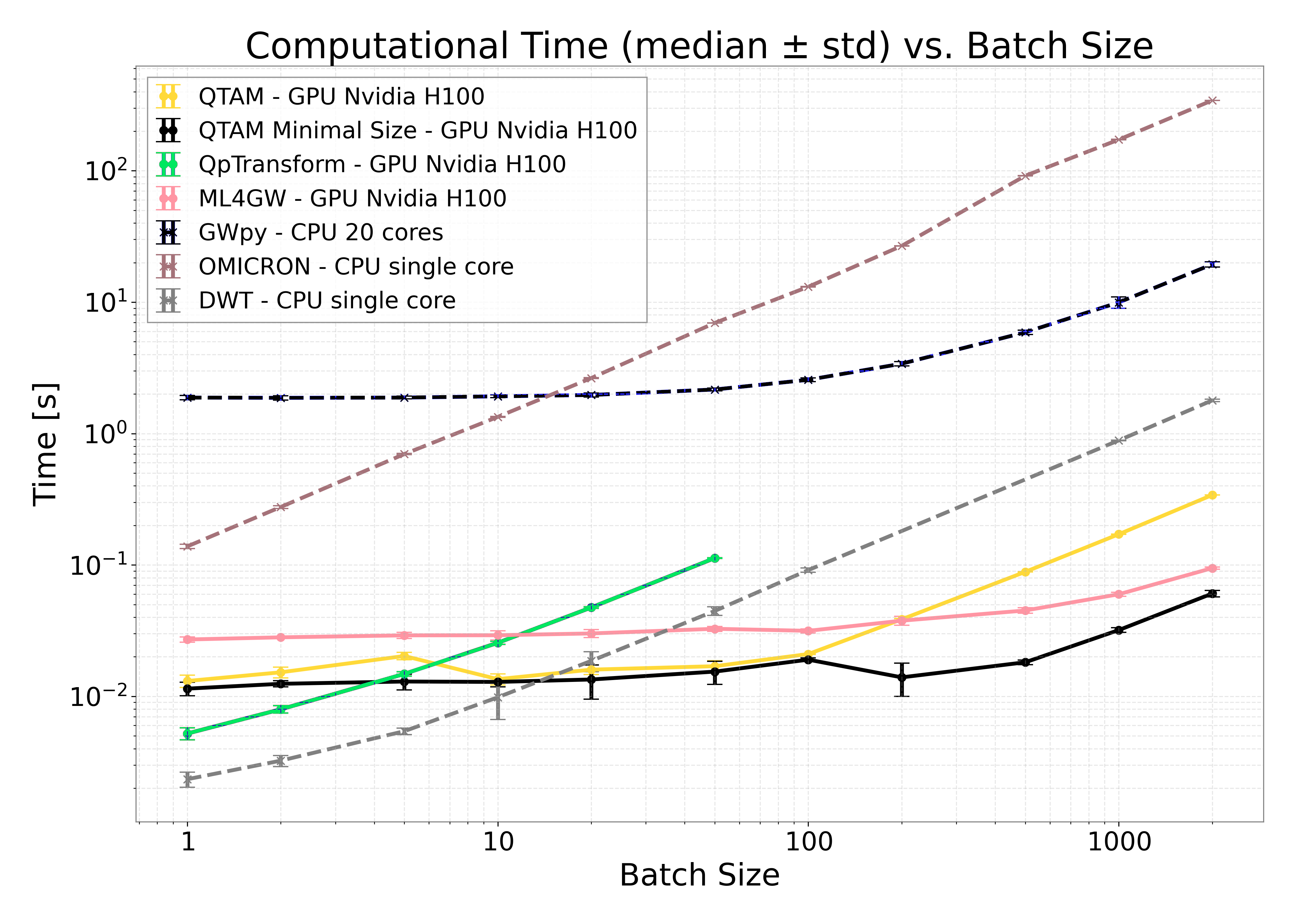}
    \caption{Computational benchmarking of CQT implementations. The plot displays the median execution time over 10 tries ($\pm$ standard deviation) as a function of input batch size on a double logarithmic scale. We compare the proposed \texttt{QTAM} algorithm (yellow) against state-of-the-art libraries: \texttt{GWpy} (blue), Wavelet Q-Transform (experimental GPU implementation) (green), \texttt{ml4gw} (pink), Omicron (brown) and the DWT wavelet library employed by \texttt{coeherent WaveBurst} (gray). Dashed lines indicate CPU execution, the number of cores is the one given by the standard settings of each software.
 Solid lines represent GPU execution performed on an NVIDIA H100. The times for the \texttt{QpTransform} were only measured up to batch size 100 because over that size the GPU memory filled up.  \texttt{QTAM} demonstrates superior scaling, maintaining sub-second latency bounds even for batch sizes of $\mathcal{O}(10^3)$.} 
    \label{fig:times}
\end{figure}

To verify \texttt{QTAM}'s suitability for low-latency search pipelines, we performed extensive benchmarking against the software described in Sec.~\ref{sec:sota}. Performance tests utilized a dataset of 6s time series sampled at 4096~Hz, containing glitches from the O3a Virgo run. We measured processing speeds on both CPU and GPU across increasing batch sizes. It was chosen to benchmark each software with its default physical parameters, such as the number of frequency tiles and the number of points sampled for each tile. This results in outputs of different sizes for different codes, which are detailed in Appendix \ref{app:performance}. We benchmarked the \texttt{QTAM} software for two possible choices of outputs: the standard-sized matrices obtained without setting any extra parameters and the minimal-sized matrices with a number of points per tile equal to  $n=\lceil N_t\cdot\Delta f_{max} \rceil +1$. This choice is made to showcase the usefulness of \texttt{QTAM} in storing the physical information in a compact form which can be computed faster than any other alternative for high batch sizes. Each software has different default settings for its usage of the available cores in the hardware, these were also left unchanged and they are specified in Fig. \ref{fig:times} and \ref{fig:times_cpu}. \texttt{Coherent WaveBurst} is natively designed as a parallel code using Message Passing Interface (MPI). For this benchmarking we chose to run the code on a single core, but the results for higher number of cores can be easily extrapolated from the ones in Fig. \ref{fig:times}. All the codes tested for the benchmaring produce CQTs, with the exception of the script producing a Discrete Wavelet Transform (DWT) transform from \texttt{Coherent WaveBurst}, as explained in Sec. \ref{sec:sota}. The computational times of the DWTs are not meant to be directly compared with the ones for the computation of CQTs, but rather they should be used as a reference for the order of magnitude of the timings required to run on the official LVK pipelines.

As shown in Fig.~\ref{fig:times}, \texttt{QTAM} maintains speeds well within low-latency bounds, even when processing batches of $\mathcal{O}(10^3)$ transforms on a GPU. The results obtained running exclusively on CPU are in Appendix \ref{app:performance}.

\texttt{QTAM} demonstrates superior speed compared to established implementations like Omicron \footnote{For the comparison with Omicron, we implemented a \texttt{C++} script utilizing Omicron's core class functions. Some parameters were adjusted from standard settings to ensure an output matrix of fixed resolution ($500 \times 1000$), matching the resolution computed by \texttt{QTAM} and the other implementations. It is important to note that this benchmarking isolates the speed of computing the CQT as complex matrices; Omicron is designed as a full transient detection pipeline, including trigger management and parallel channel handling, which are outside the scope of this specific speed test.} used upstream current low-latency pipelines. These results position \texttt{QTAM} as a highly efficient building block upon which to construct next-generation analysis infrastructures. 
While the detailed characterization of such complete pipelines is outside the scope of this work, we present a targeted proof-of-concept in Sec.~\ref{sec:denoise} to demonstrate the algorithm's potential in tackling critical challenges such as signal de-noising and disentanglement by reproducing the clustering and inversion logic used by established burst searches \cite{cWB, burst}.

\section{Application: signal de-noising and disentangling} \label{sec:denoise}
In this section, we present the application of \texttt{QTAM} to signal de-noising and the disentanglement of overlapping transient events. As third-generation observatories like ET and CE become operational, the temporal overlap of GW signals is expected to become commonplace, introducing significant biases in parameter estimation \cite{Samajdar_2021,Sathyaprakash_2012}. A similar challenge is already present in the current detector network, where transient noise artifacts (glitches) act as a persistent source of systematic uncertainty. We propose to address both issues by exploiting the sparsity of GW signals in the TF domain. Since \texttt{QTAM} allows to resolve signals that are temporally overlapping but distinct in their frequency evolution, it offers a robust method to mitigate systematic biases in both future high-rate regimes and current noise-dominated environments.

\begin{figure}[h!] 
    \centering
    \includegraphics[width=0.99\linewidth]{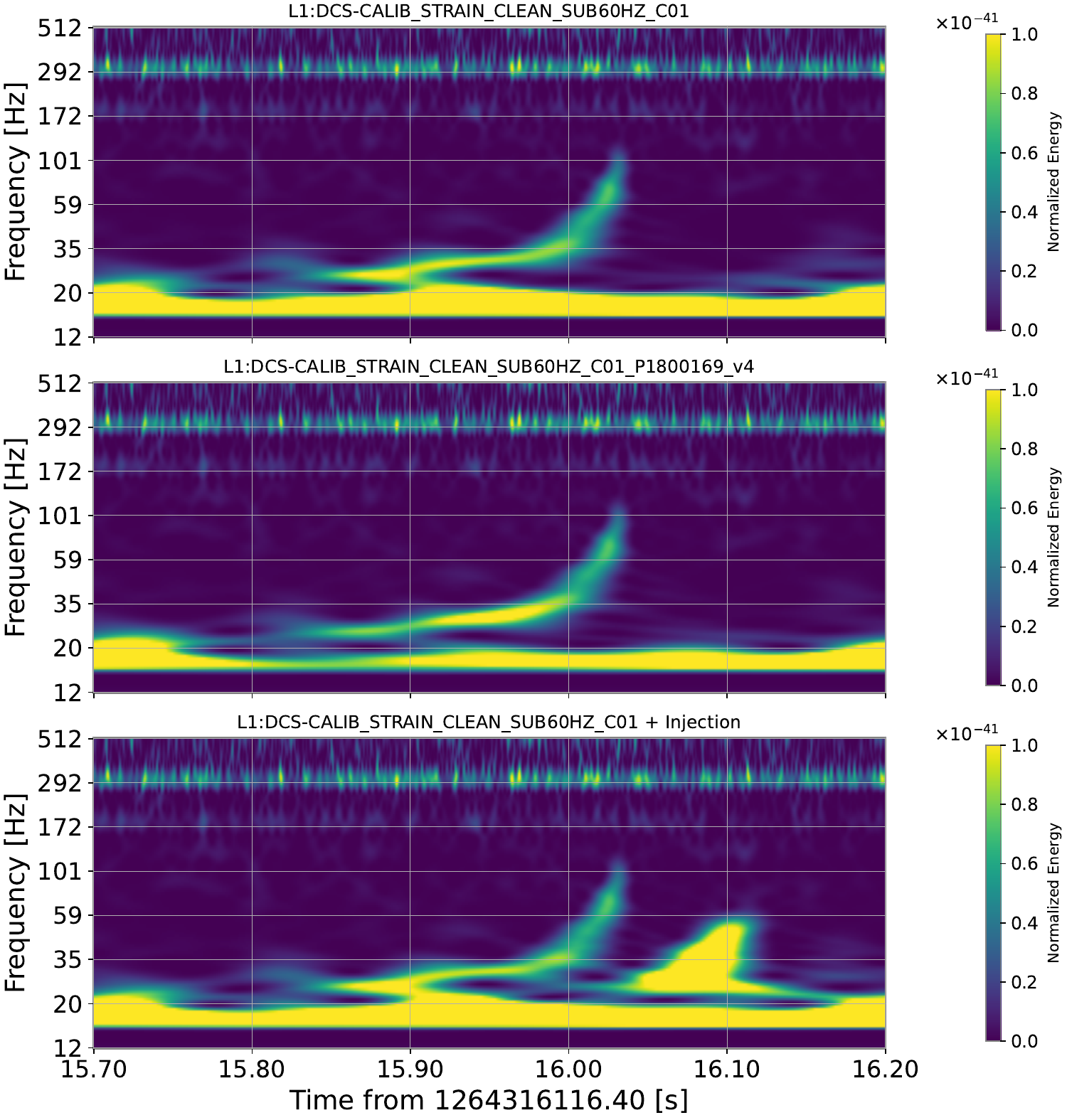}
    \caption{TF representation ($Q=12$) of the GW200129\_065458 event in L1. 
    The figure displays Q-transform spectrograms of the detector strain data surrounding the merger time. \textit{Top panel:} The original L1 data stream \textit{Middle panel:} The corresponding glitch-subtracted data with standard LVK methods.
    \textit{Bottom panel} The L1 data stream with the injected wave. For full parameters see Appendix \ref{app:denoising}.}
    \label{fig:gw200129}
\end{figure}

To evaluate the efficacy of our approach, we selected the event GW200129\_065458 as a test case. GW200129\_065458 is a highly significant Binary Black Hole (BBH) merger detected during the O3b observing run \cite{GWTC3, Hannam22}. This event is notable for its potential evidence of strong-field spin-induced orbital precession \cite{Hannam22} and large recoil velocities \cite{Varma22}, but primarily for the data quality challenges associated with it. The signal coincided with broadband non-Gaussian noise in the LIGO Livingston (L1) detector, attributed to the 45 MHz electro-optic modulator system \cite{Macas23, Davis22} as show in Fig. \ref{fig:gw200129}. 

The dependence of the inferred physical parameters on the specific glitch mitigation technique used (e.g., \texttt{gwsubtract}, \texttt{BayesWave}) remains a subject of debate \cite{Gupte24, Payne22}, making this event an ideal benchmark for novel de-noising techniques. To extend this analysis to the high-rate regime expected in future observing runs, we further augmented the dataset by injecting a synthetic BBH signal temporally overlapping with the real GW200129\_065458 event as shown in Fig. \ref{fig:gw200129}.

\begin{figure}[h!]
    \centering
    \includegraphics[width=0.99\linewidth]{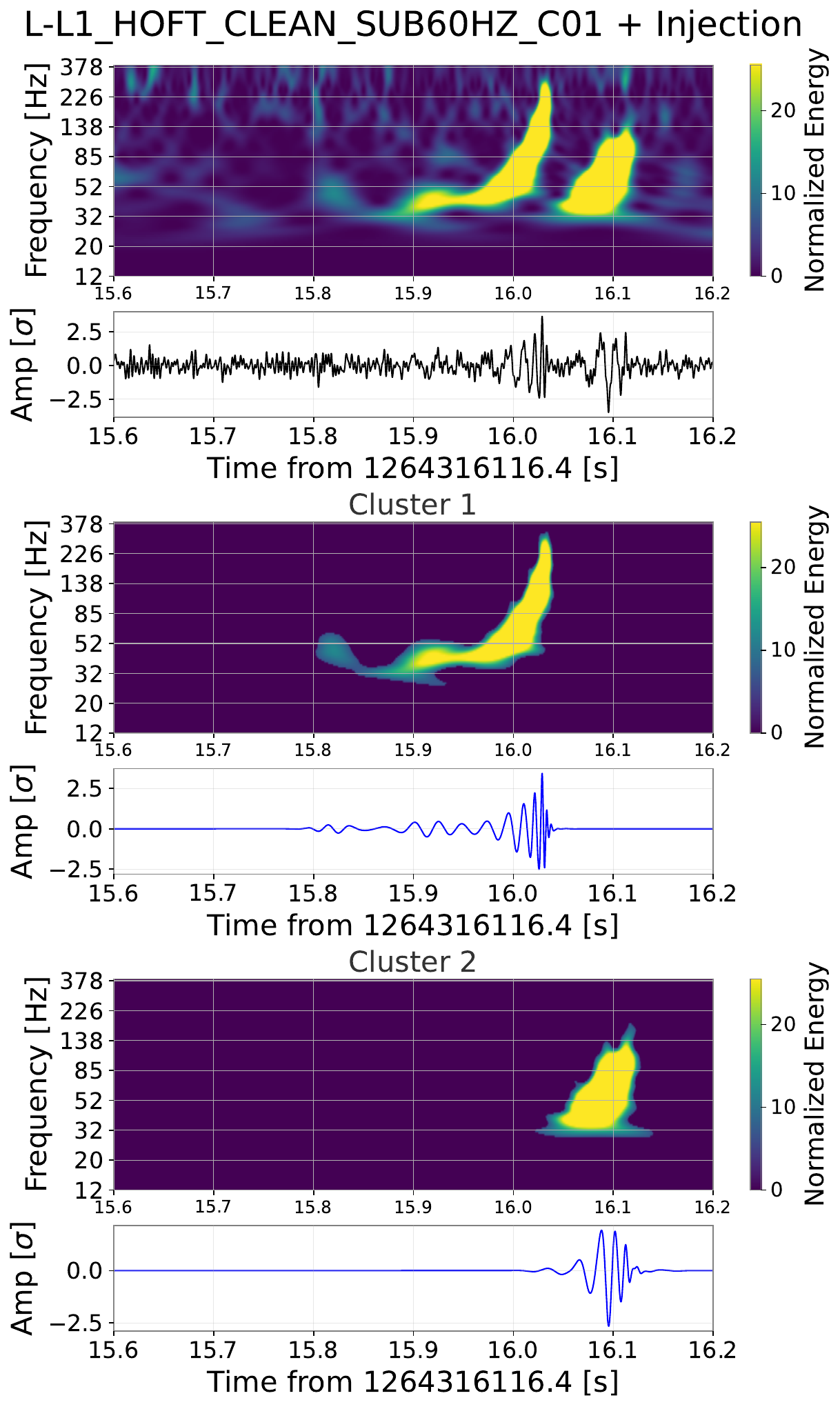}
     \caption{Illustrative signal recovery via attention clustering on whitened data. \textit{Top panel:} The full Q-transform spectrogram and corresponding inverted time-series of strain data containing a synthetic injection. \textit{Middle and Bottom panels:} Decomposition of the signal into distinct attention clusters (Cluster 1 and Cluster 2). By masking the complex CQT coefficients to these isolated TF regions and applying the inverse transform, the individual time-domain components are reconstructed, validating the extraction of coherent waveform features from background noise.}
    \label{fig:cluster_inversion}
\end{figure}

\begin{figure}[h!]
    \centering
    \includegraphics[width=0.99\linewidth]{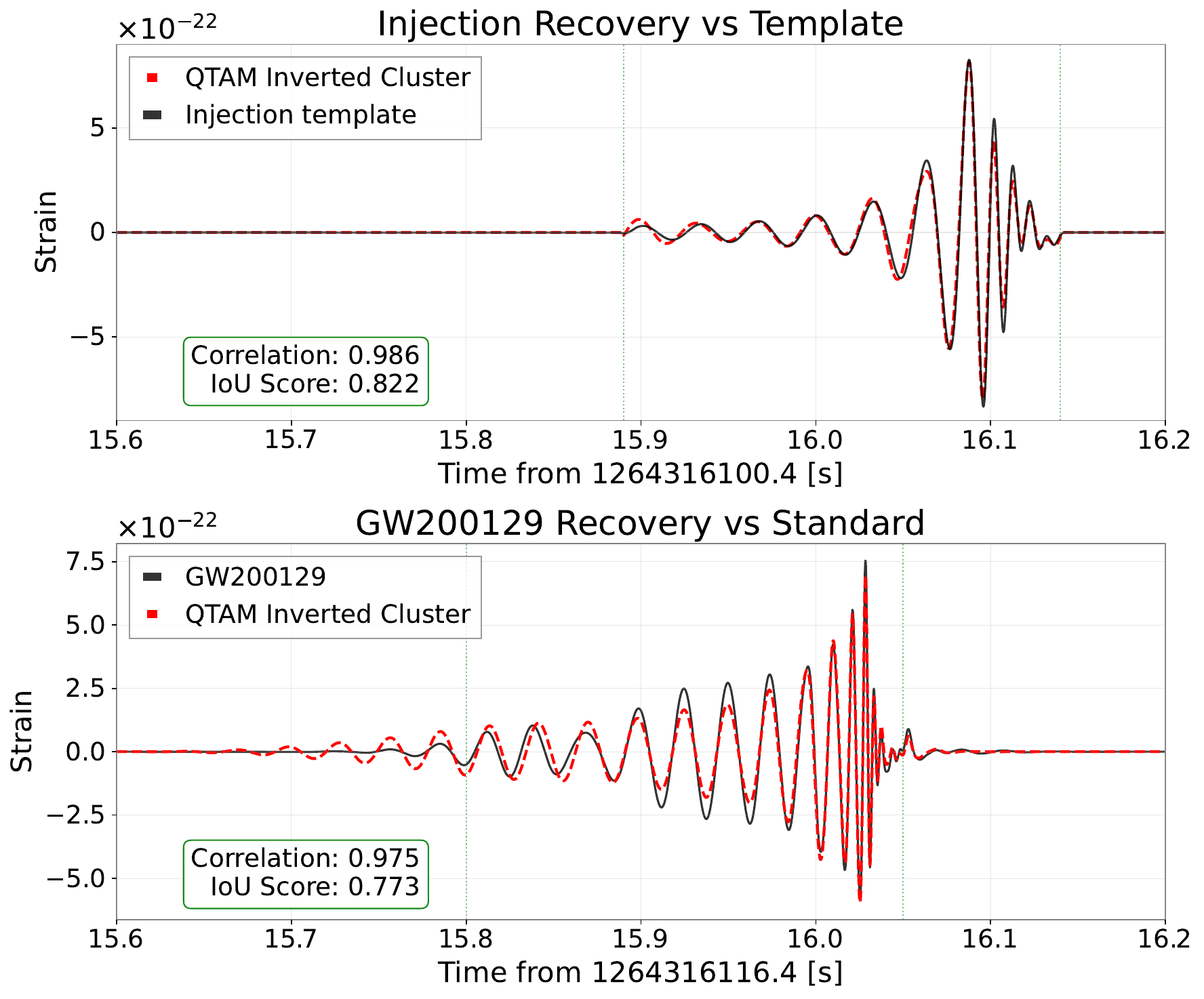}
    \caption{Validation of \texttt{QTAM} clustering inversion signal recovery.
    \textit{Top:} The solid black line represents the ground truth injected waveform, while the dashed red line shows the signal recovered by applying the inverse Q-transform to the identified attention cluster.
    \textit{Bottom:} Reconstruction of the real gravitational wave event GW200129\_065458. The black line denotes the standard pre-processed strain data, overlaid with the \texttt{QTAM} cluster reconstruction (red dashed) after the deglitching process discussed in the main text. The discrepancies between the two waveforms are expected and due to the different denosing procedures applied.
    Green vertical dotted lines delimit the temporal window used to calculate the Pearson correlation coefficient ($r$) and the Intersection over Union (IoU) score, displayed in the insets. The high metric values demonstrate that the attention mask accurately captures the phase and amplitude evolution of the chirp.}
    \label{fig:Injection_recovery}
\end{figure}

We applied \texttt{QTAM} to this composite dataset to isolate the astrophysical signals from both the glitch and one another. Adopting a reconstruction strategy similar to the Coherent WaveBurst (cWB) pipeline \cite{cWB}, we implemented a clustering algorithm to identify contiguous energetic regions based on a tunable normalized threshold and minimum area. For each identified cluster, we generated a binary mask applied to the complex CQT, which was subsequently inverted to recover the time-domain waveform as shown in Fig. \ref{fig:cluster_inversion}. 
Such a clustering-inversion process was performed iteratively, alternating between low ( $Q\leq 6$ and high ($Q\geq 12$) Q-factors to disentangle events in time and frequency respectively. This iterative strategy exploits the fundamental nature of the CQT as an overcomplete frame rather than an orthonormal basis. While the DWT ensures independence between coefficients via a fixed, critically sampled grid, it is inherently limited by a static resolution trade-off and a lack of shift-invariance. In contrast, the CQT frame sacrifices orthogonality for redundancy, allowing for arbitrary frequency centering and tunable TF resolution. By coupling this redundant representation with non-linear clustering, our method functions as an adaptive sparse approximation: low-Q passes allow for the precise excision of transient features in the time domain, while subsequent high-Q passes resolve the remaining spectral features in the frequency domain, effectively circumventing the resolution limits of a single-pass linear projection.\\
The injected wave was successfully recovered with a Pearson correlation coefficient of $r=98.6\%$ (despite being injeceted onto a region characterixed by the presence of glitches in the $20\text{-}40\text{Hz}$ range) as shown in Fig. \ref{fig:Injection_recovery}.

To further quantify the impact of \texttt{QTAM} on astrophysical inference, we performed a full Bayesian parameter estimation analysis. We utilized the \texttt{bilby} inference library \cite{Ashton19} with the \texttt{SEOBNRv5\_ROM} waveform approximant \cite{PhysRevD.108.124037} and a minimum frequency cutoff of $f_{\text{min}} = 30$ Hz. We compared four distinct data treatments: the standard uncleaned frames (STD), the state-of-the-art BayesWave-cleaned frames (BW), the full \texttt{QTAM} reconstruction derived from the raw data, and a hybrid dataset where \texttt{QTAM} was used solely as a filter to suppress residual background noise in the $20\text{--}40$ Hz band of the BayesWave cleaned frames (BW+\texttt{QTAM}).

\begin{figure}[t]
    \centering
    \includegraphics[width=0.99\linewidth]{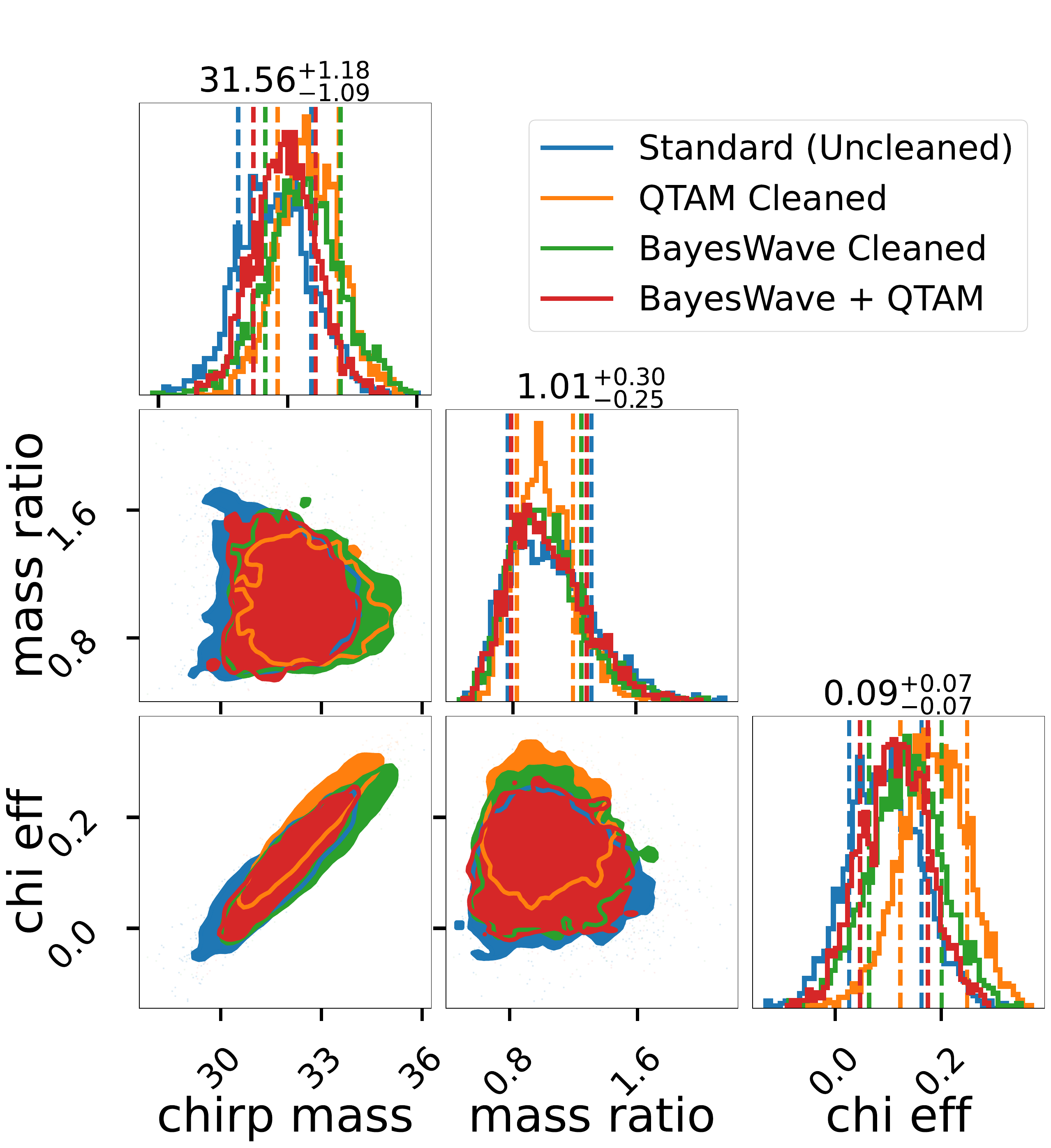}
    \caption{
    Posterior distributions for intrinsic parameters of the injected binary black hole system: chirp mass $\mathcal{M}$, mass ratio $q$, and effective inspiral spin $\chi_{\mathrm{eff}}$. 
    }
    \label{fig:corner_intrinsic}
\end{figure}

\begin{figure}[t]
    \centering
    \includegraphics[width=0.99\linewidth]{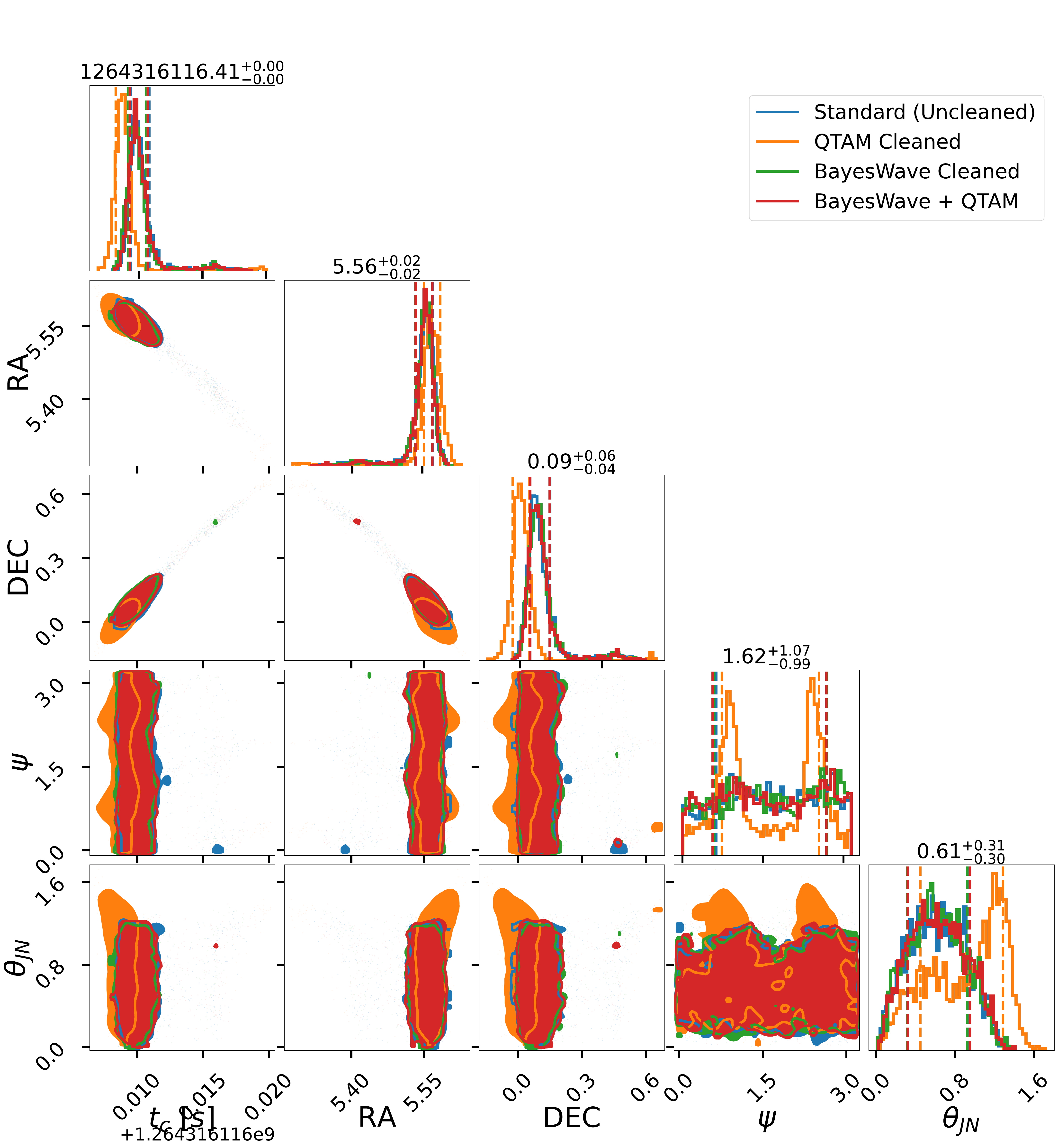}
    \caption{
    Posterior distributions for extrinsic parameters: geocentric time $t_c$, right ascension $\alpha$, declination $\delta$, inclination angle $\theta_{JN}$, and polarization $\psi$. 
    }
    \label{fig:corner_extrinsic}
\end{figure}

The resulting posterior distributions, shown in~\cref{fig:corner_extrinsic,fig:corner_intrinsic}, exhibit a clear dependence on the adopted data treatment.

For intrinsic parameters such as the chirp mass, mass ratio, and effective spin, all configurations yield broadly consistent posteriors with large mutual overlaps, indicating that the source's core physical properties are robust to the choice of cleaning strategy. In contrast, parameters that rely heavily on phase coherence, such as coalescence time and sky localization, exhibit more pronounced differences. Notably, the \texttt{QTAM}-only reconstruction produces posteriors that are visibly narrower but shifted relative to standard baselines. The nature of these shifts is best clarified by the hybrid BW+\texttt{QTAM} configuration. In this case, we observe a consistent reduction in posterior width of $\sim 10-40\% $ compared to BW alone, yet without the parameter deviations seen in the standalone \texttt{QTAM} analysis. This suggests that the method itself is not intrinsically biased, a conclusion further reinforced by the injection study in Fig.~\ref{fig:Injection_recovery}, where \texttt{QTAM} preserves signal fidelity provided sufficient TF separation is available. The shifts observed in the standalone reconstruction likely reflect the difficulty of isolating the signal where it overlaps with the glitch. In such regions, the clustering mask may inadvertently clip parts of the astrophysical signal, causing the small phase distortions we observe. A secondary possibility is that \texttt{QTAM} removes residual noise components that standard methods leave behind, thereby modifying the effective noise realization and sharpening the likelihood surface.

While further systematic validation is required to weigh these competing hypotheses and fully validate \texttt{QTAM} impact on denosising, the broader utility of the method is clear. These results highlight \texttt{QTAM}'s strong potential for signal purification, but also suggest that a simple iterative clustering strategy is not a silver bullet for unbiased inference. Instead, the optimal approach appears to be integration: as the BW+\texttt{QTAM} configuration demonstrates, combining \texttt{QTAM} with signal-agnostic cleaning allows for tighter constraints without compromising accuracy. The development of such hybrid strategies, capable of navigating complex noise environments with high fidelity, will be the focus of future work.
\section{Conclusions.} 
We have introduced \texttt{QTAM}, a framework designed to resolve the fundamental trade-off in TF analysis between the computational efficiency of critically sampled transforms and the shift-invariance of overcomplete representations. By demodulating the Q-transform coefficients to baseband, our method achieves lossless decimation that retains the phase coherence and tunable resolution essential for deep learning, while reducing data redundancy to the Nyquist-Shannon sampling limit defined by the signal bandwidth. This enables fully invertible, high-fidelity processing within the strict latency bounds of online alert pipelines.\\

The application to real detector data highlights the potential of these invertible spectrograms for signal denoising and disentanglement, while requireing furhter systemtic studies. However, our analysis reveals that a simple strategy of iterative masking is insufficient on its own to guarantee unbiased parameter estimation in complex noise environments. While effective at isolating features, naive clustering can inadvertently modify the signal structure or the noise realization in ways that bias inference.\\

Instead, the optimal path forward lies in integrating these precise morphological constraints into signal-agnostic reconstruction frameworks. By combining the ability of \texttt{QTAM} to resolve distinct evolutionary tracks with the statistical robustness of established inference pipelines, it is possible to tighten constraints on overlapping signals without compromising accuracy. This hybrid capability will be indispensable for the era of third-generation observatories, providing the necessary foundation to disentangle the high density of concurrent events expected in the next generation of gravitational wave astronomy.

\section*{Code availability}
The \texttt{QTAM} package is publicly available as a Python library and can be installed via pip. The complete source code, documentation, and usage examples are hosted on GitHub at \url{https://github.com/dottormale/Qtransform_torch/tree/main/QTAM}. Users can install the package with the command \texttt{pip install qtam}, which automatically handles all dependencies including PyTorch and NumPy.
\section*{Acknowledgments}
The authors thank W. Benoit, E. Cuoco, M. Coughlin, F. Di Renzo, E. Milotti, F. Robinet, A. Virtuoso for helpful discussions. LA, FS and AR acknowledge financial support from the interTwin project funded by the European Union Horizon Europe Programme – Grant Agreement number 101058386. JL acknowledges support from the Italian Ministry of University and
Research (MUR) via the PRIN 2022ZHYFA2, GRavitational wavEform models for coalescing compAct binaries with eccenTricity (GREAT).\\
This research has made use of data or software obtained from the Gravitational Wave Open Science Center (http://gwosc.org), a service of LIGO Laboratory, the LIGO Scientific Collaboration, the Virgo Collaboration, and KAGRA. This material is based upon work supported by NSF's LIGO Laboratory which is a major facility fully funded by the National Science Foundation. LIGO Laboratory and Advanced LIGO are funded by the United States National Science Foundation (NSF) as well as the Science and Technology Facilities Council (STFC) of the United Kingdom, the Max–Planck–Society (MPS), and the State of Niedersachsen/Germany for support of the construction of Advanced LIGO and construction and operation of the GEO600 detector. Additional support for Advanced LIGO was provided by the Australian Research Council. Virgo is funded, through the European Gravitational Observatory (EGO), by the French Centre National de Recherche Scientifique (CNRS), the Italian Istituto Nazionale di Fisica Nucleare (INFN) and the Dutch Nikhef, with contributions by institutions from Belgium, Germany, Greece, Hungary, Ireland, Japan, Monaco, Poland, Portugal, Spain. KAGRA. is supported by Ministry of Education, Culture, Sports, Science and Technology (MEXT), Japan Society for the Promotion of Science (JSPS) in Japan; National Research Foundation (NRF) and Ministry of Science and ICT (MSIT) in Korea; Academia Sinica (AS) and National Science and Technology Council (NSTC) in Taiwan. 

\appendix
\section{QTAM implementation details} \label{app:implementation}

The \texttt{QTAM} software is architected as a modular \texttt{PyTorch} \texttt{nn.Module}, designed to maximize throughput via GPU tensor operations. The codebase is organized into two primary components: the core transform logic and the multi-Q extension.

To support diverse analysis requirements, \texttt{QTAM} implements distinct window functions vectorized natively in \texttt{PyTorch}. Because the implementation leverages the Convolution Theorem, these functions are generated directly as discrete spectral filters $\tilde{W}[k]$. 
For a filter centered at index $k_c$ with a bandwidth $\Delta k_f$, the support $\mathcal{K}_{k_c}$ is defined as in Eq.~(\ref{eq:support}) of the main text.
The standard window for GW analysis is the \textit{Bi-square} window, selected for its minimal spectral leakage:
\begin{equation}
    \tilde{W}[k] = \left( 1 - \left( \frac{k - k_c}{\Delta k_f} \right)^2 \right)^2
\end{equation}
Alternative options include the \textit{Hann} window, corresponding to a cosine bell in the frequency domain:
\begin{equation}
    \tilde{W}[k] = \frac{1}{2} \left[ 1 + \cos\left( \pi \frac{k - k_c}{\Delta k_f} \right) \right]
\end{equation}
We also utilize the \textit{Kaiser} window, which allows control over the side-lobe attenuation via the shape parameter $\beta$. It is implemented using \texttt{PyTorch}'s native zeroth-order modified Bessel function of the first kind ($I_0$):
\begin{equation}
    \tilde{W}[k] = \frac{I_0\left( \beta \sqrt{1 - \left( \frac{k - k_c}{\Delta k_f} \right)^2} \right)}{I_0(\beta)}
\end{equation}
For adjustable tapering, we implement the \textit{Tukey} window, controlled by the parameter $\alpha_T$ representing the fraction of the window inside the cosine taper:
\begin{widetext}
\begin{equation}
    \tilde{W}[k] = \begin{cases} 
    1 & \text{if } |k - k_c| \le \Delta k_f (1 - \alpha_T) \\
    \frac{1}{2} \left[1 + \cos\left(\pi \frac{\frac{|k - k_c|}{\Delta k_f} - (1 - \alpha_T)}{\alpha_T}\right)\right] & \text{if } \Delta k_f (1 - \alpha_T) < |k - k_c| \le \Delta k_f \\
    0 & \text{otherwise}
    \end{cases}
\end{equation}
\end{widetext}
Finally, for extremely smooth transitions, we implement the \textit{Planck-taper} window, utilizing the Planck distribution function defined by the boundary width fraction $\alpha_P$. The window is given by:
\begin{equation}
    \tilde{W}[k] = \frac{1}{e^{Z(k)} + 1}
\end{equation} 
where the auxiliary function $Z(k)$ governs the transition regions $\Delta k_f(1-\alpha_P) < |k - k_c| < \Delta k_f$:
\begin{equation}
    Z(k) = \frac{\alpha_P}{1 - \frac{|k - k_c|}{\Delta k_f}} + \frac{\alpha_P}{\frac{|k - k_c|}{\Delta k_f} - (1 - \alpha_P)}
\end{equation}
Additionally, to ensure strict element-wise invertibility even in the stopband (as discussed in the main text), any of the window functions above can be initialized with a non-zero regularization term $\epsilon \ll 1$ outside the spectral support $\mathcal{K}_{k_c}$.

\begin{figure}[h!]
    \centering
    \includegraphics[width=0.99\linewidth]{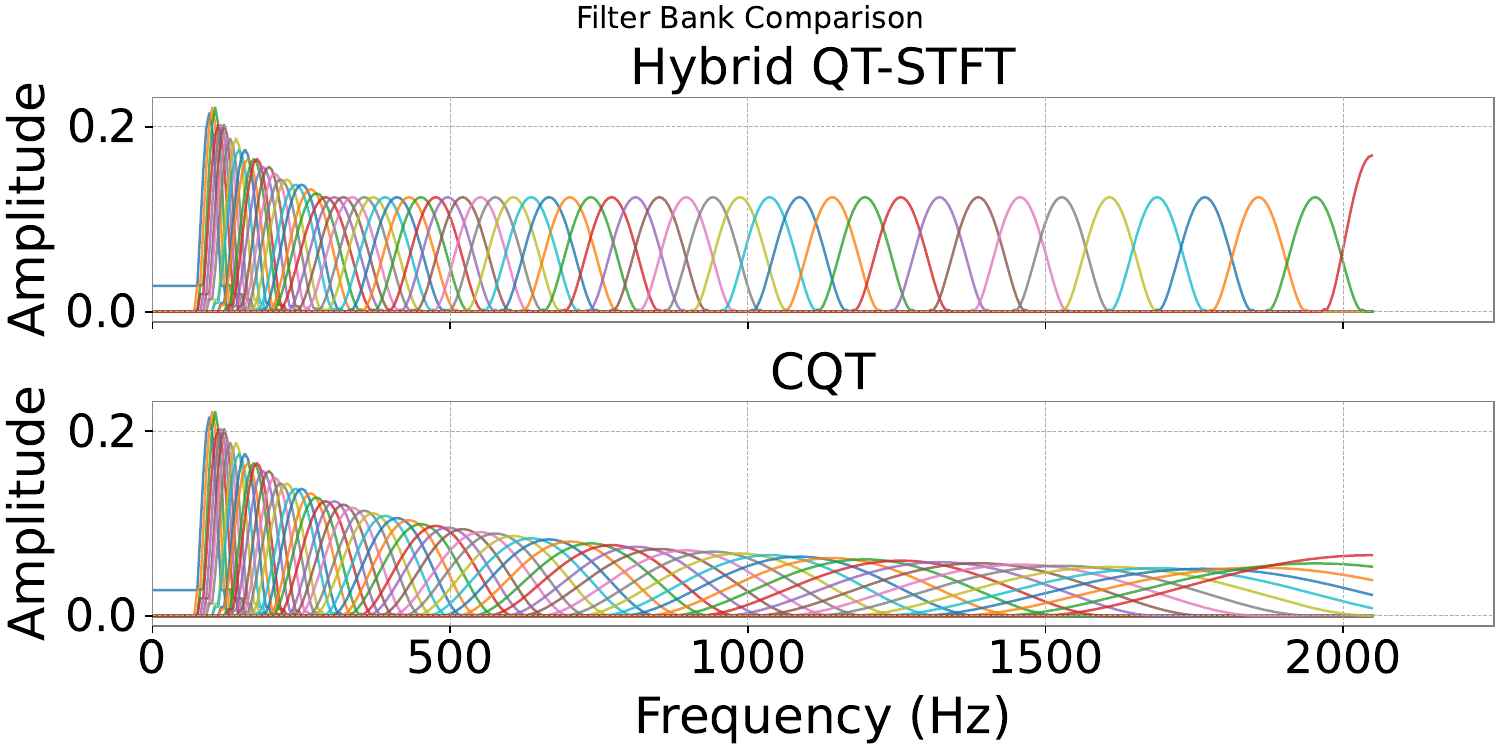}
    \caption{Comparison of Filter Bank Topologies. The plot illustrates the frequency response of the \texttt{QTAM} window basis in the frequency domain. \textit{Top:} The hybrid mode employed by \texttt{QTAM}. At low frequencies, the bandwidths vary geometrically (CQT behavior); at high frequencies, the bandwidth clamping ($L_{max}$) forces the filters to maintain a constant width (STFT behavior), limiting the maximum sampling rate required for downsampling. \textit{Bottom:} A pure CQT formulation without constraints. Note how the filters at the high spectral edge become extremely wide, necessitating a very high sampling rate to avoid aliasing during demodulation.}
    \label{fig:filterbank}
\end{figure}

\begin{figure}[t!]
    \centering
    \includegraphics[width=0.99\linewidth]{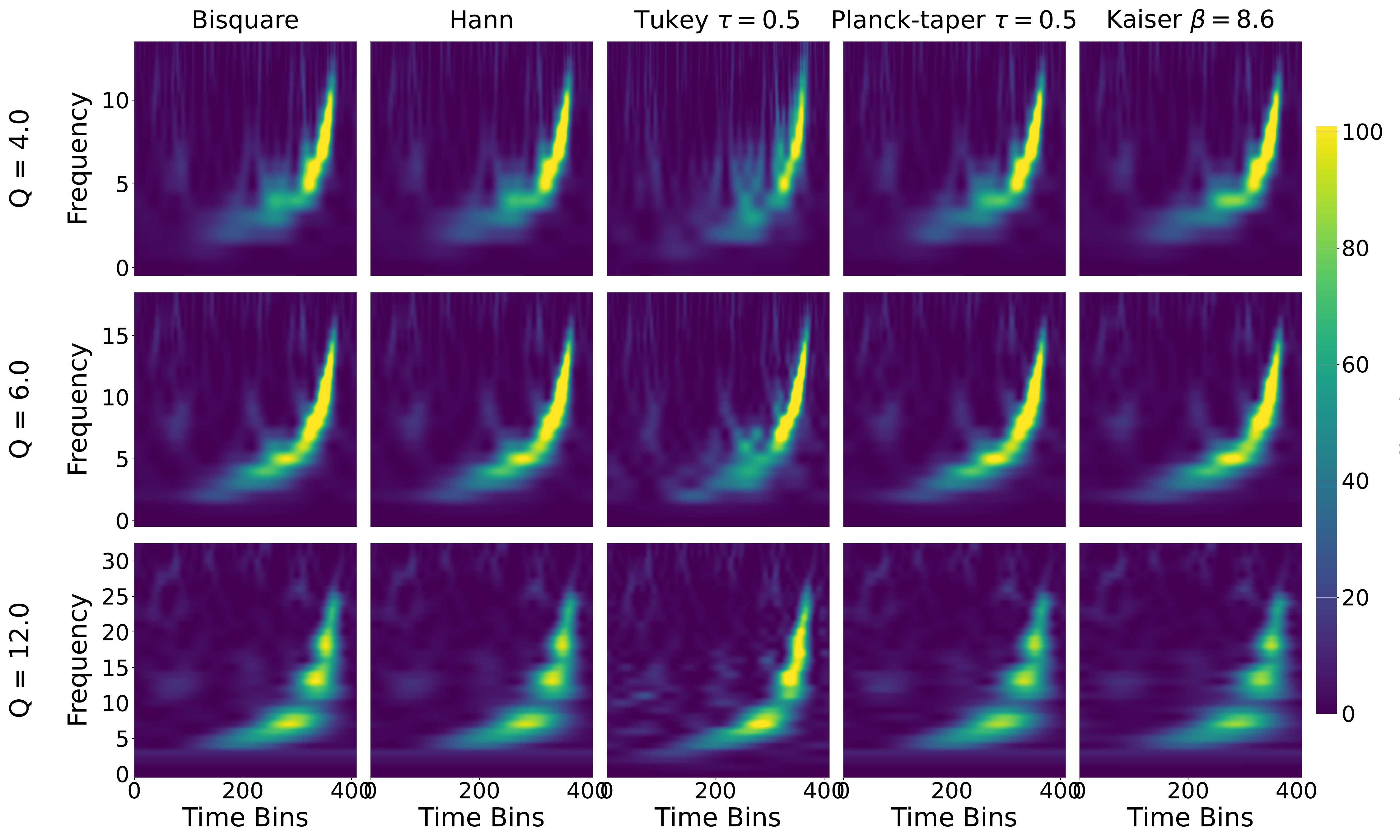} 
    \caption{Multi-Parameter Q-Scan of GW150914 (H1). The figure displays energy spectrograms obtained by simultaneously scanning across varying Quality factors ($Q$) and window configurations. The rows correspond to increasing $Q$ values ($Q=4, 8, 12$), illustrating the variable trade-off between time and frequency resolution. The columns demonstrate the effect of different window functions and shape parameters (e.g., Bisquare, Tukey with varying $\tau$, Kaiser with varying $\beta$). This vectorized approach facilitates the rapid identification of the optimal representation maximizing signal visibility.}
    \label{fig:qscan}
\end{figure}

The spectral grid can be defined either manually, by passing a specific list of target frequencies, or automatically. In the automatic mode, the user defines the frequency range $[f_{min}, f_{max}]$ and the resolution in bins per octave $B$, generating a geometric progression $f_k = f_{min} \cdot 2^{k/B}$. 
Once the grid is established, the algorithm determines the window length $L_{k_c}$ (corresponding to the filter's spectral support $\mathcal{K}_{k_c}$ in frequency bins) for each central frequency. In this frequency-domain implementation, the ideal Constant-Q bandwidth scales linearly with frequency: $L_k^{ideal} \propto f_k \cdot (N_t/Q)$. To prevent excessive bandwidths at high frequencies (which would dictate a prohibitively high sampling rate for lossless downsampling) the software enforces a global constraint \texttt{max\_window\_size}. This constraint can be set manually by the user or determined automatically. In the latter case, the algorithm calculates the minimal bandwidth required to ensure invertibility (no spectral gaps) by computing the maximum spacing between consecutive center frequencies, $\Delta f_{max} = \max(f_{max} - f_{k_{max}-1})$, and setting the cap to cover this gap:
\begin{equation}
    L_{max} = 2 \left\lfloor \frac{\lceil N_t \cdot \Delta f_{max} \rceil}{2} \right\rfloor + 1
\end{equation}
The spectral window length for each filter is then derived via the clamping logic:
\begin{equation}
    L_k = \min( L_k^{ideal}, L_{max} )
\end{equation}
This mechanism forces \texttt{QTAM} to transition seamlessly between a pure Constant-Q Transform at low frequencies (where bandwidths are narrow, $L_k^{ideal} < L_{max}$) and a Short-Time Fourier Transform (STFT) at high frequencies (where bandwidth is clamped to $L_{max}$). For the typical range of GW signals ($\mathcal{O}(10^2)$~Hz), the transform retains its pure CQT character, preserving the variable TF resolution required to fully capture transient morphologies. At higher frequencies, the hybrid topology enforces structural constraints on window duration. This yields a dual advantage: it allows the demodulation downsampling to be significantly more aggressive by constraining the global maximum bandwidth, and ensures that the filter bank covers the full spectral support of the signal while minimizing computational cost, as shown in Fig. \ref{fig:filterbank}.
\\
\\
\texttt{QTAM} leverages the Convolution Theorem to perform filtering efficiently in the frequency domain. The process begins with a single Real-to-Complex FFT (RFFT) of the input batch $X$, yielding the frequency series $\tilde{X}(f)$. The filter bank is implemented as a set of pre-computed, normalized spectral windows $\tilde{W}_k(f)$ stored in the \texttt{full\_window} buffer of each \texttt{QTile} module. These windows are pre-shifted during initialization to be centered exactly on their respective target frequencies $f_k$. For each target frequency, the filtering operation is a fast element-wise multiplication:
\begin{equation}
    Y_k(f) = \tilde{X}(f) \cdot \tilde{W}_k(f)
\end{equation}
This isolates the spectral content of the signal in the band of interest. The amplitude demodulation step is integrated directly into this computation sequence. After filtering, the resulting spectrum $Y_k(f)$ is circularly shifted (rolled) by $-f_k$ to move the signal content to DC (baseband). Once aligned, the spectrum is cropped to retain only the coefficients within the target bandwidth. This allows the subsequent Inverse FFT to be executed on a drastically reduced support ($T_{out} \ll N$), contrasting with standard pipelines that typically compute a full-length transform before applying destructive interpolation. Consequently, this approach ensures linearity and exact invertibility while scaling computational cost with the compressed output size rather than the input duration.\\
\\
This core logic is extended to analyze the signal across a high-dimensional parameter space simultaneously. The algorithm manages a bank of transforms, scanning not only across varying Quality factors $Q$ (determined via a user-provided list or a geometric progression) but also across diverse windowing functions and their associated shape parameters (e.g., taper width $\tau$, Kaiser $\beta$). To ensure computational efficiency, this sweep is fully vectorized: a composite filter bank concatenates all requested configurations, enabling the simultaneous computation of the entire parameter space in a single execution step.\\
For each resolution plane, the algorithm computes the spectrogram and evaluates a signal quality metric specified by the user (e.g., total energy or peak pixel intensity). The implementation supports two primary operational modes. In the \textit{full scan} mode, it returns the complete set of spectrograms for all requested $Q$ values and window configurations, preserving the full high-dimensional volume for detailed inspection as illustrated in Fig. \ref{fig:qscan}. 
In the \textit{optimal selection} mode, the algorithm dynamically selects the single best plane based on the energy criterion. A key feature of this mode is the ability to perform the maximization independently for each signal in the batch. This allows the pipeline to automatically adapt the TF resolution and windowing strategy to the specific morphology of individual events within a heterogeneous dataset, yielding a composite output tensor constructed from the optimal planes
A defining feature of the \texttt{QTAM} implementation is its full integration with the \texttt{PyTorch} computation graph. By inheriting from the library's base classes and relying exclusively on differentiable operations (FFT/IFFT, complex multiplication, grid sampling), the entire pipeline is compatible with automatic differentiation (\texttt{autograd}). This allows gradients to flow backward from the output spectrogram to the input time series, enabling \texttt{QTAM} to be embedded directly into custom loss functions (e.g., minimizing spectral distance) for generative tasks or denoising autoencoders. Furthermore, the implementation treats the batch dimension $N_{batch}$ independently, allowing simultaneous processing of thousands of signals on a GPU, ensuring high throughput when used as a pre-processing layer within end-to-end Neural Network architectures.

\section{QTAM performance and profiling}\label{app:performance}

The benchmarking of all the software packages described in this work were conducted on a machine equipped with 384 cores and a NVIDIA H100 graphics processing unit with 80GB of memory. The test dataset comprised time series containing Scattered Light glitches from the O3 observing run, stored in a PyTorch tensor of dimensions $8740 \times 24576$ . In the case of \texttt{GWpy}, the inputs were converted into \texttt{TimeSeries} objetcs, and for \texttt{Omicron} into \texttt{Numpy} arrays.  For experiments involving variable batch sizes, each batch was constructed by randomly sampling a fixed number of one-dimensional time series from this tensor.

The profiling of the \texttt{QTAM} software focused on two primary aspects: computational speed and memory usage. The procedure consisted of executing a dedicated Python script that invoked the principal functions implemented in the \texttt{QTAM} software.  Regarding memory consumption, particular attention was devoted to ensuring that memory requirements remained within acceptable limits throughout all computational stages, thereby guaranteeing that the software can operate on standard machines without risk of memory overflow. We monitored not only GPU memory usage during the computation of the transform, but also memory consumption during the transfer of data from CPU to GPU. Our analysis reveals no anomalous or problematic behavior, with memory usage remaining modest throughout all operations. The standard \texttt{QTAM} outputs have dimensions $62 \times 12289$, while the \texttt{QTAMs} of minimal size have dimensions $62 \times 669$.

\begin{figure}[h!] 
    \centering
    \includegraphics[width=0.99\linewidth]{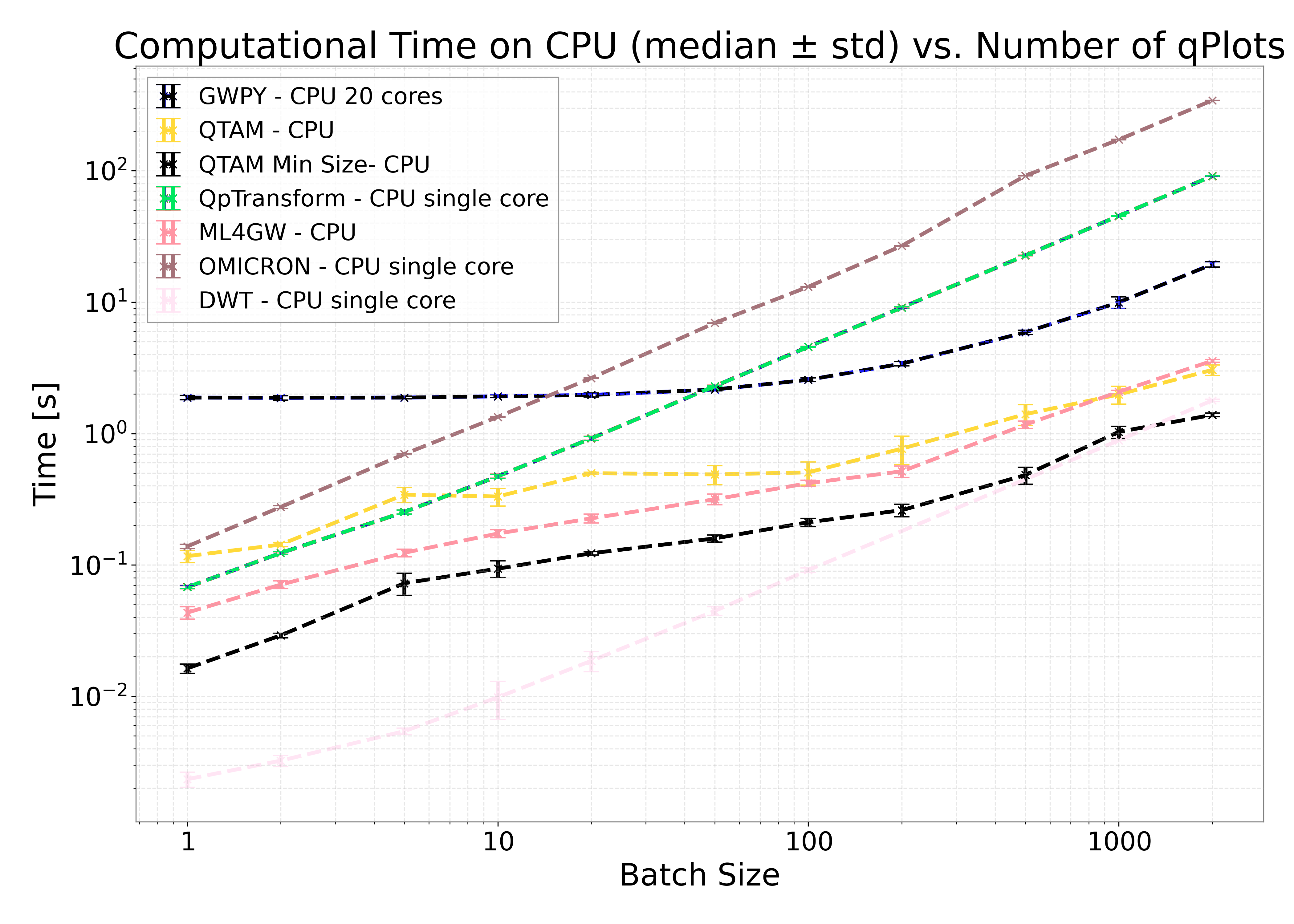}
    \caption{Computational benchmarking of CQT implementations running on CPU. The plot displays the median execution time over 10 tries ($\pm$ standard deviation) as a function of input batch size on a double logarithmic scale. The \texttt{Omicron} and \texttt{ml4gw} lines are the same as showin in Fig \ref{fig:times}. The number of CPU cores is specified for the codes which use a fixed number of them; both \texttt{ml4gw} and \texttt{QTAM} dinamically exploit the ones available on the machine, therefore no single number is specified. For the codes which run on a single core, the computational time scales linearly with the batch sizes as expected.
    }
    \label{fig:times_cpu}
\end{figure}

The performance shown in Fig. \ref{fig:times_cpu} for \texttt{GWpy} is achieved exploiting 20 CPU cores, which is the standard setting for the package. The output CQTs have standard dimensions $500 \times 1000$.

The benchmarking of the QpTransform was performed using the code in \cite{UnoaccasoRepo}. The original code is designed to run on GPU using the CuPy library. In order to test the code on CPU, we adapted the algorithm by substituting the CuPy object with equivalent NumPy ones, specifically the Just-In-Time Raw Kernels employed in the GPU version have been substituted with Numba kernels which target the CPU. The basic functions employed in the code do not perform operations in batch but rather sequentially, therefore the computational time scales linearly with the batch dimension, as expected. The output CQTs have dimensions $ 62 \times 24576$, since the number of points for each tile always has to be equal to the sample rate (4096~Hz) multiplied by the length of the signal (6~s).
For the GPU run, the data was organised inside Dask Arrays to manage their parallel processing. Dask also ensures CUDA kernel compilation at runtime, effectively eliminating a systematic overhead of approximately 1 s and isolating the primary computational performance.

The \texttt{ml4gw} software uses the same algorithm as \texttt{GWpy}, but the code is implemented in \texttt{PyTorch}. This change enables a faster performance in CPU but shows its real advantages when running on GPU, allowing for the processing of tensors of high batch size and much better scaling. For the benchmark we set the output CQTs to have the same dimensions as the ones of \texttt{GWpy}: $500 \times 1000$.

For the comparison with the Omicron software, we wrote a \texttt{C++} script with Omicron's elementary class functions. Some parameters were set as different from Omicron's standard, so to ensure that we would have as output a matrix of fixed resolution $500 \times 1000$, the same as the one computed by \texttt{ml4gw}. It is important to specify that the benchmarking was done exclusively by comparing the speed of computation of CQT as complex matrices; software like Omicron is not designed to perform only this limited task, but to carry out a larger analysis for a transient event detection, including trigger management and handling of multiple channels in parallel. While limited in scope, this comparison is sufficient for the purpose of our tool, which is not to have an alternative detection pipeline, but to have a new package which is very fast and is suitable for machine learning applications and parameter estimation, as discussed in Sec. \ref{sec:denoise}.

The computational times on CPU for the different implementations of the CQT are shown in Fig.\ref{fig:times_cpu}.

\section{Invertibility of the discrete Q-transform and preservation under QTAM demodulation}
\label{app:invertibility}

The invertibility of the discrete Q-transform is most naturally established by interpreting the transform as a finite frame expansion in frequency space. Although the continuous Q-transform does not generally satisfy the admissibility condition required by continuous wavelet inversion theory \cite{Daubechies1992}, its discrete implementation belongs to the finite-dimensional setting of frame theory, where exact reconstruction depends only on completeness and boundedness of the analysis window family \cite{grochenig2001foundations, holighaus2013framework}.

Let $x[n]\in\mathbb{C}^{N}$ denote a finite discrete signal, with discrete Fourier transform:
\begin{equation}
X[k]=\sum_{n=0}^{N-1}x[n]e^{-i2\pi kn/N}
\end{equation}
and inverse transform:
\begin{equation}
x[n]=\frac{1}{N}\sum_{k=0}^{N-1}X[k]e^{i2\pi kn/N}
\end{equation}

For each frequency channel $k$, the discrete Q-transform coefficient is obtained by multiplying the spectrum by a frequency window $\tilde W_k[f]$ and transforming back to the time index:
\begin{equation}
T_k[\tau]
=
\frac{1}{N}\sum_{f=0}^{N-1}X[f]\tilde W_k[f]e^{i2\pi f\tau/N}
\end{equation}
This coefficient can be written as an inner product with the atom:
\begin{equation}
g_{k,\tau}[n]
=
\frac{1}{N}\sum_{f=0}^{N-1}\tilde W_k[f]e^{i2\pi f(n-\tau)/N}
\end{equation}
so that:
\begin{equation}
T_k[\tau]=\langle x,g_{k,\tau}\rangle 
\end{equation}
The discrete Q-transform therefore defines an analysis operator generated by the family $\{g_{k,\tau}\}$.\\
Unlike an orthonormal basis, the family $\{g_{k,\tau}\}$ need not be linearly independent. In practical Q-transform constructions the number of atoms generally exceeds the signal dimension, so the representation is redundant and coefficient sequences are not unique. Exact reconstruction nevertheless remains possible provided that $\{g_{k,\tau}\}$ forms a frame for $\mathbb{C}^{N}$, meaning that there exist constants $A,B>0$ such that
\begin{equation}
A\|x\|^2
\le
\sum_{k,\tau}|\langle x,g_{k,\tau}\rangle|^2
\le
B\|x\|^2   
\end{equation}
for every $x\in\mathbb{C}^{N}$. The middle quantity represents the total coefficient energy, so the frame inequality states that coefficient energy and signal energy remain equivalent up to fixed multiplicative bounds.

Let us now define the analysis operator
\begin{equation}
T:\mathbb{C}^{N}\rightarrow\mathbb{C}^{M},
\qquad
(Tx)_{k,\tau}=\langle x,g_{k,\tau}\rangle
\end{equation}
where $M$ denotes the total number of coefficients. Its norm satisfies
\begin{equation}
\|Tx\|^2
=
\sum_{k,\tau}|\langle x,g_{k,\tau}\rangle|^2
\end{equation}

so the frame inequality becomes:
\begin{equation}
A\|x\|^2
\le
\|Tx\|^2
\le
B\|x\|^2    
\end{equation}

The lower bound implies injectivity of $T$: if $Tx=0$, then $\|Tx\|^2=0$, hence
\begin{equation}
A\|x\|^2\le 0   
\end{equation}
which forces $x=0$ because $A>0$. Thus no nonzero signal can produce a vanishing coefficient family.\\
The adjoint operator
\begin{equation}
T^*:\mathbb{C}^{M}\rightarrow\mathbb{C}^{N}
\end{equation}
acts as synthesis:
\begin{equation}
T^*c
=
\sum_{k,\tau} c_{k,\tau} g_{k,\tau}  
\end{equation}

Composing analysis and synthesis yields the frame operator:
\begin{equation}
S=T^*T
\end{equation}
which acts explicitly as:
\begin{equation}
Sx
=
\sum_{k,\tau}\langle x,g_{k,\tau}\rangle g_{k,\tau}   
\end{equation}

To establish invertibility, consider the quadratic form associated with $S$:
\begin{equation}
\langle Sx,x\rangle
=
\left\langle
\sum_{k,\tau}\langle x,g_{k,\tau}\rangle g_{k,\tau},x
\right\rangle  
\end{equation}
Using linearity of the inner product:
\begin{equation}
\langle Sx,x\rangle
=
\sum_{k,\tau}
\langle x,g_{k,\tau}\rangle
\langle g_{k,\tau},x\rangle  
\end{equation}
\\
By Hermitian symmetry,
\begin{equation}
\langle g_{k,\tau},x\rangle
=
\overline{\langle x,g_{k,\tau}\rangle}
\end{equation}
so that:
\begin{equation}
\langle Sx,x\rangle
=
\sum_{k,\tau}
|\langle x,g_{k,\tau}\rangle|^2    
\end{equation}

Substituting into the frame inequality gives:
\begin{equation}
A\|x\|^2
\le
\langle Sx,x\rangle
\le
B\|x\|^2    
\end{equation}
\\
Hence $S$ is bounded, self-adjoint, and positive definite. Self-adjointness follows from:
\begin{equation}
    \langle Sx, y \rangle = \langle T^*Tx, y \rangle = \langle Tx, Ty \rangle = \langle x, T^*Ty \rangle = \langle x, Sy \rangle
\end{equation}
Because $S$ is positive definite in finite dimension, all eigenvalues are strictly positive. If:
\begin{equation}
Su=\lambda u   
\end{equation}
then:
\begin{equation}
\langle Su,u\rangle
=
\lambda\|u\|^2  
\end{equation}
and the frame inequality gives:
\begin{equation}
A\|u\|^2
\le
\lambda\|u\|^2
\le
B\|u\|^2   
\end{equation}

Since $u\neq 0$:
\begin{equation}
A\le \lambda \le B
\end{equation}

All eigenvalues therefore lie strictly away from zero, so $S$ is invertible.\\
The reconstruction formula follows once the explicit Fourier structure of $S$ is computed. Starting from
\begin{equation}
Sx[n]
=
\sum_{k,\tau} T_k[\tau]\, g_{k,\tau}[n]
\end{equation}
substitute the atom definition:
\begin{equation}
Sx[n]
=
\sum_{k,\tau}
T_k[\tau]
\left(
\frac{1}{N}
\sum_{f=0}^{N-1}
\tilde W_k[f]e^{i2\pi f(n-\tau)/N}
\right)    
\end{equation}
Reordering the sums gives:
\begin{equation}
Sx[n]
=
\frac{1}{N}
\sum_k
\sum_{f=0}^{N-1}
\tilde W_k[f]e^{i2\pi fn/N}
\sum_{\tau=0}^{N-1}
T_k[\tau]e^{-i2\pi f\tau/N}
\end{equation}

The inner sum is the discrete Fourier transform of the coefficient sequence:
\begin{equation}
\tilde T_k[f]
=
\sum_{\tau=0}^{N-1}
T_k[\tau]e^{-i2\pi f\tau/N}   
\end{equation}
Substituting the definition of $T_k[\tau]$,
\begin{equation}
T_k[\tau]
=
\frac{1}{N}
\sum_{\ell=0}^{N-1}
X[\ell]\tilde W_k[\ell]e^{i2\pi \ell\tau/N}
\end{equation}
yields:
\begin{equation}
\tilde T_k[f]
=
\sum_{\tau=0}^{N-1}
\left(
\frac{1}{N}
\sum_{\ell=0}^{N-1}
X[\ell]\tilde W_k[\ell]e^{i2\pi \ell\tau/N}
\right)e^{-i2\pi f\tau/N}    
\end{equation}
Exchanging summation order:
\begin{equation}
\tilde T_k[f]
=
\frac{1}{N}
\sum_{\ell=0}^{N-1}
X[\ell]\tilde W_k[\ell]
\sum_{\tau=0}^{N-1}
e^{i2\pi(\ell-f)\tau/N} 
\end{equation}
Using discrete Fourier orthogonality
\begin{equation}
\sum_{\tau=0}^{N-1}
e^{i2\pi(\ell-f)\tau/N}
=
N\delta_{\ell f} 
\end{equation}
gives:
\begin{equation}
\tilde T_k[f]
=
X[f]\tilde W_k[f]  
\end{equation}
Substituting back into the expression for $Sx[n]$,
\begin{equation}
Sx[n]
=
\frac{1}{N}
\sum_k
\sum_{f=0}^{N-1}
X[f]\tilde W_k[f]\tilde W_k^*[f]e^{i2\pi fn/N}
\end{equation}
The channel sum can be collected:
\begin{equation}
Sx[n]
=
\frac{1}{N}
\sum_{f=0}^{N-1}
X[f]
\left(
\sum_k |\tilde W_k[f]|^2
\right)
e^{i2\pi fn/N}  
\end{equation}
Define the aggregate spectral coverage
\begin{equation}
P[f]
=
\sum_k |\tilde W_k[f]|^2 
\end{equation}
Then:
\begin{equation}
Sx[n]
=
\frac{1}{N}
\sum_{f=0}^{N-1}
P[f]X[f]e^{i2\pi fn/N}   
\end{equation}
Taking the discrete Fourier transform gives
\begin{equation}
(SX)[f]
=
P[f]X[f]  
\end{equation}
Thus the frame operator is diagonal in the Fourier basis: every spectral component is multiplied independently by the total local window energy contributed by all overlapping channels. The frame bounds are therefore equivalent to pointwise bounds
\begin{equation}
A\le P[f]\le B
\end{equation}
Invertibility now reduces to strict positivity:
\begin{equation}
P[f]>0
\qquad
\forall f
\end{equation}
If $P[f_0]=0$ for some frequency $f_0$, then all windows vanish there and the component $X[f_0]$ is lost. Conversely,
\begin{equation}
S^{-1}X[f]
=
\frac{X[f]}{P[f]}
\end{equation}
Applying $S^{-1}$ yields the canonical dual-frame reconstruction:
\begin{equation}
x[n]
=
\mathcal{F}^{-1}
\left\{
\frac{\sum_k \tilde T_k[f]\tilde W_k^*[f]}
{\sum_k |\tilde W_k[f]|^2}
\right\}   
\end{equation}
Substituting
\begin{equation}
\tilde T_k[f]=X[f]\tilde W_k[f]    
\end{equation}
gives:
\begin{equation}
\hat X[f]
=
\frac{\sum_k X[f]\tilde W_k[f]\tilde W_k^*[f]}
{P[f]}
=
X[f]   
\end{equation}
hence:
\begin{equation}
\hat x[n]=x[n] 
\end{equation}
This proves exact reconstruction of the discrete Q-transform whenever the aggregate spectral coverage remains strictly positive.

The QTAM construction preserves this invertibility because demodulation acts independently in each channel as a unitary spectral translation followed by bandwidth-adapted decimation. Let $\tilde T_k[f]$ denote the original tile centered at frequency $f_k$. QTAM applies
\begin{equation}
\tilde Y_k[f]
=
\tilde T_k[f+f_k]
\end{equation}
which shifts the spectral support to baseband. Since spectral translation corresponds to multiplication by a complex exponential in time, this operation is unitary and preserves inner products by Parseval's identity \cite{oppenheim1999signals}.

After translation, the effective support is bounded by the local bandwidth $\Delta f_k$, so temporal sampling only needs to satisfy the Nyquist condition associated with that bandwidth rather than the original carrier frequency. The inverse transform length changes from $N_{\mathrm{in}}$ to $N_{\mathrm{out}}$, with normalization
\begin{equation}
Y_k[n]
=
\sqrt{\frac{N_{\mathrm{out}}}{N_{\mathrm{in}}}}
\,\mathcal{F}^{-1}\{\tilde Y_k[f]\}  
\end{equation}
This normalization preserves energy:
\begin{equation}
\sum_n |T_k[n]|^2
=
\sum_n |Y_k[n]|^2  
\end{equation}
During synthesis, the inverse operation restores the original tile:
\begin{equation}
\tilde T_k[f]
=
\sqrt{\frac{N_{\mathrm{in}}}{N_{\mathrm{out}}}}
\,\tilde Y_k[f-f_k]  
\end{equation}

Thus the original frame coefficients are recovered exactly before synthesis, and the reconstruction formula remains unchanged:
\begin{equation}
x[n]
=
\mathcal{F}^{-1}
\left\{
\frac{\sum_k \tilde T_k[f]\tilde W_k^*[f]}
{\sum_k |\tilde W_k[f]|^2}
\right\}.  
\end{equation}
QTAM therefore leaves the frame operator unchanged:
\begin{equation}
P[f]
=
\sum_k |\tilde W_k[f]|^2
\end{equation}
so the same frame bounds remain valid:
\begin{equation}
0<A\le P[f]\le B<\infty
\end{equation}

Exact reconstruction follows with machine precision provided that identical spectral shifts and normalization factors are applied during synthesis.

This also clarifies why QTAM differs fundamentally from interpolation-based compression: interpolation introduces non-unitary projections that alter the frame operator and destroy dual-frame consistency, whereas QTAM performs only invertible unitary conjugations together with bandwidth-consistent resampling, preserving the exact inverse map.

\section{GW200129\_065458 Denoising} \label{app:denoising}
In this appendix, we detail the data selection, pre-processing, and synthetic injection methodology used to evaluate the efficacy of the \texttt{QTAM} as a denoising and signal disentangling tool on the GW200129\_065458 event.

The event GW200129\_065458, detected during the third observing run (O3b) of the LIGO-Virgo network \cite{GWTC3}, serves as the primary test case for this study due to the presence of a high-amplitude, broadband noise transient in the LIGO Livingston (L1) detector coincident with the merger. To evaluate the performance of our pipeline against established mitigation strategies, we retrieved 32 seconds of strain data centered on the event GPS time ($t_0 = 1264316116.4$) for the LIGO Hanford (H1), LIGO Livingston (L1), and Virgo (V1) observatories. For L1, we analyzed two distinct data versions: the standard calibrated strain (\texttt{C01} frames) containing the unmitigated glitch, and a specialized clean data frame (\texttt{L1:DCS-CALIB\_STRAIN\_CLEAN\_SUB60HZ\_C01\_P1800169\_v4}). In the latter, the 45 MHz electro-optic modulator noise was removed using linear subtraction based on auxiliary witness sensors \cite{Davis22, Driggers19}. 

For the spectral whitening, we employed the standard on-source power spectral density (PSD) estimation method utilized in LVK parameter estimation analyses. The PSDs were computed directly from the 32-second data segments using the \texttt{BayesWave} algorithm \cite{Cornish15, Littenberg15}. 

To simulate the signal overlap expected in next-generation detector networks, we augmented the real data with a synthetic Binary Black Hole (BBH) injection. The data were decimated to a sampling rate of 4096 Hz and a synthetic waveform was added to the stream. Generated using the \texttt{bilby} library \cite{Ashton19} with the \texttt{IPhenomD} approximant, the injection represents a heavy stellar-mass system ($m_1 = 98 M_{\odot}$, $m_2 = 78 M_{\odot}$) at a luminosity distance of 2500 Mpc. The signal was injected with a time shift of $+0.085$ s relative to the GW200129\_065458 merger, resulting in significant time overlap. All parameters are reported in Table \ref{tab:injection_params}.
\begin{table}[h!]
    \centering
    \caption{Summary of injection parameters and waveform generator settings.}
    \label{tab:injection_params}
    \begin{tabular}{l c r}
        \toprule
        \textbf{Parameter} & \textbf{Symbol} & \textbf{Value} \\
        \midrule
        \multicolumn{3}{l}{\textit{Source Properties}} \\
        Primary Mass & $m_1$ & $98.0 \, M_\odot$ \\
        Secondary Mass & $m_2$ & $78.0 \, M_\odot$ \\
        Dimensionless Spin 1 & $a_1$ & $0.0$ \\
        Dimensionless Spin 2 & $a_2$ & $0.0$ \\
        Tilt 1 & $\theta_1$ & $0.0$ \\
        Tilt 2 & $\theta_2$ & $0.0$ \\
        Azimuthal Angle & $\phi_{12}$ & $0.0$ \\
        Precession Phase & $\phi_{JL}$ & $0.0$ \\
        \midrule
        \multicolumn{3}{l}{\textit{Extrinsic Properties}} \\
        Luminosity Distance & $d_L$ & $2500.0 \, \text{Mpc}$ \\
        Right Ascension & $\alpha$ & $5.95 \, \text{rad}$ \\
        Declination & $\delta$ & $-0.81 \, \text{rad}$ \\
        Inclination Angle & $\theta_{JN}$ & $0.0 \, \text{rad}$ \\
        Polarization & $\psi$ & $0.0 \, \text{rad}$ \\
        Coalescence Phase & $\phi_c$ & $0.0 \, \text{rad}$ \\
        Geocentric Time & $t_c$ & 1264316116.485 s\\
        \midrule
        \multicolumn{3}{l}{\textit{Waveform Configuration}} \\
        Approximant & -- & \texttt{IMRPhenomD} \\
        Minimum Frequency & $f_{\text{min}}$ & $30.0 \, \text{Hz}$ \\
        Source Model & -- & \texttt{lal\_binary\_black\_hole} \\
        \bottomrule
    \end{tabular}
\end{table}

We perform Bayesian parameter estimation using the \texttt{bilby} framework with the \texttt{SEOBNRv5ROM} waveform approximant, adopting a minimum frequency cutoff of $f_{\mathrm{min}} = 30 \mathrm{Hz}$. The analysis is carried out on a $4\mathrm{s}$ data segment sampled at $4096 \mathrm{Hz}$, centered around the event time. We assume standard stationary Gaussian noise characterized by pre-computed power spectral densities for each detector. To reduce parameter degeneracies and isolate the impact of the data conditioning, we restrict the analysis to aligned-spin binary black hole configurations. The detector-frame component masses are sampled independently from uniform priors in the range $m_{1,2} \in [20, 60],M_{\odot}$, while the dimensionless spin components along the orbital angular momentum are drawn from uniform distributions $\chi_{1,2} \in [-0.99, 0.99]$. The luminosity distance is sampled from a uniform-in-comoving-volume prior between $200$ and $2000,\mathrm{Mpc}$, and the geocentric coalescence time is assigned a uniform prior within $\pm 0.1,\mathrm{s}$ of the trigger time. Phase and distance marginalization are employed to improve sampling efficiency. Posterior sampling is performed using the \texttt{dynesty} nested sampler with $1000$ live points.
To assess the impact of the different denoising strategies on parameter estimation, we compare the posterior distributions obtained from the four data treatments described in the main text: Standard (STD), QTAM, BayesWave (BW), and the combined BW+QTAM configuration.

\begin{figure*}[t]
    \centering
    \includegraphics[width=\textwidth]{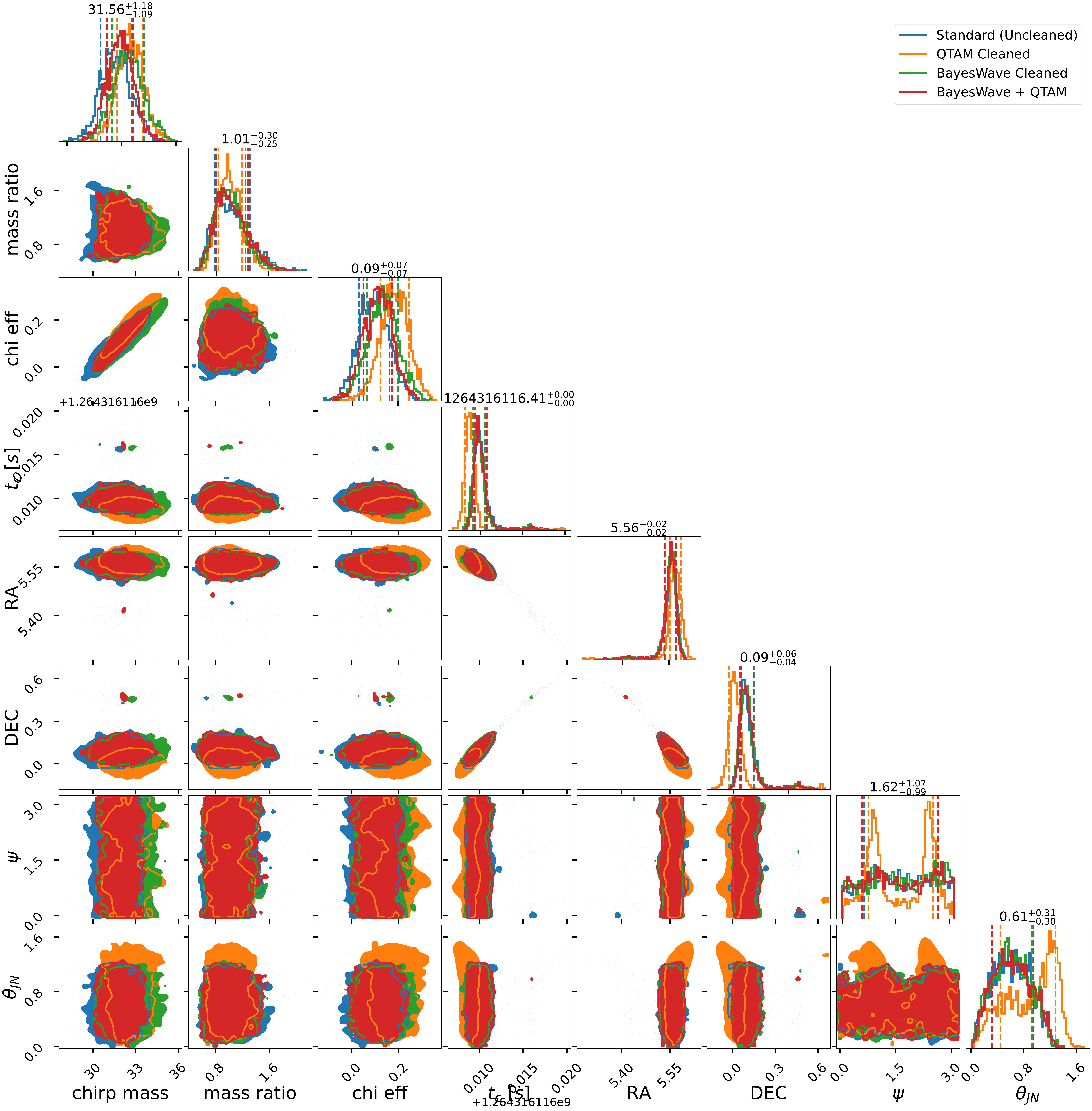}
    \caption{
    Full posterior distributions for all sampled parameters for the injected signal. 
    This figure complements Figs.~\ref{fig:corner_intrinsic} and \ref{fig:corner_extrinsic} by displaying the complete parameter space. 
    The BW+QTAM configuration consistently yields tighter constraints while remaining compatible with the injected values. 
    The QTAM-only reconstruction shows visible shifts in several parameters, consistent with partial signal–glitch overlap affecting the reconstruction.
    }
    \label{fig:corner_full}
\end{figure*}

Figure~\ref{fig:corner_full} shows the full corner plot for all recovered parameters, while the main text presents reduced corner plots separating intrinsic and extrinsic parameters for clarity. The full plot is included here for completeness and to allow inspection of parameter correlations.

To quantify differences between posterior distributions, we evaluate both one-dimensional and two-dimensional metrics. For the marginal distributions, we compute the ratio of posterior widths and the overlap between distributions. For the joint distributions, we compute the ratio of credible-region areas and the containment fraction between posteriors.

Table~\ref{tab:1d_summary} summarizes the comparison between BW and BW+QTAM for the most relevant parameters. 
The two-dimensional structure of the posteriors is examined in Table~\ref{tab:2d_summary}. 
Finally, Table~\ref{tab:cross_config} compares all configurations relative to the standard analysis. 

\begin{table}[t]
\centering
\caption{
Comparison of 1D posterior distributions between BayesWave (BW) and the combined BW+QTAM configuration.
Width ratios quantify changes in posterior precision (values $<1$ indicate tighter constraints), while the overlap measures distributional consistency.
}
\label{tab:1d_summary}
\begin{tabular}{lcc}
\toprule
Parameter & Width Ratio & Overlap \\
\midrule
$\mathcal{M}$        & 0.78 & 0.96 \\
$q$                  & 0.99 & 1.00 \\
$\chi_{\mathrm{eff}}$& 0.90 & 0.99 \\
$t_c$                & 1.22 & 0.99 \\
RA                   & 1.24 & 1.00 \\
Dec                  & 1.25 & 1.00 \\
\bottomrule
\end{tabular}
\end{table}

\begin{table}[t]
\centering
\caption{
Comparison of 2D posterior structure between BayesWave (BW) and BW+QTAM.
Area ratios quantify changes in joint posterior volume (values $<1$ indicate tighter constraints), while containment measures the overlap of credible regions.
}
\label{tab:2d_summary}
\begin{tabular}{lcc}
\toprule
Parameter Pair & Area Ratio & Containment \\
\midrule
$\mathcal{M}$ -- $q$              & 0.82 & 0.52 \\
$\mathcal{M}$ -- $\chi_{\mathrm{eff}}$ & 0.64 & 0.75 \\
$q$ -- $\chi_{\mathrm{eff}}$      & 0.92 & 0.49 \\
$\mathcal{M}$ -- $t_c$            & 0.85 & 0.60 \\
$\chi_{\mathrm{eff}}$ -- $t_c$    & 0.95 & 0.60 \\
RA -- Dec                         & 1.03 & 0.80 \\
\bottomrule
\end{tabular}
\end{table}

\begin{table}[t]
\centering
\caption{
Relative posterior widths for different configurations, normalized to the standard analysis (STD).
Values $<1$ indicate improved precision.
}
\label{tab:cross_config}
\begin{tabular}{lcccc}
\toprule
Parameter & STD & QTAM & BW & BW+QTAM \\
\midrule
$\mathcal{M}$        & 1.00 & 0.85 & 0.93 & 0.78 \\
$q$                  & 1.00 & 0.71 & 0.91 & 0.90 \\
$\chi_{\mathrm{eff}}$& 1.00 & 0.93 & 1.03 & 0.93 \\
$t_c$                & 1.00 & 0.52 & 0.94 & 1.15 \\
RA                   & 1.00 & 0.63 & 0.86 & 1.07 \\
Dec                  & 1.00 & 0.53 & 0.84 & 1.05 \\
\bottomrule
\end{tabular}
\end{table}

\newpage


\clearpage 

\pagenumbering{arabic}
\setcounter{page}{1}

\bibliographystyle{unsrt}
\def\apjl{Astroph.~J.~Lett.}
\def\LVCAbbott{{LIGO Scientific Collaboration and Virgo Collaboration (B.P.~Abbott)}}
\bibliography{references}

@book{Daubechies1992,
    author = {Daubechies, I.},
    title = {Ten lectures on wavelets},
    publisher = {SIAM},
    year = {1992},
}

@inproceedings{velasco2011constructing,
  title={Constructing an invertible constant-Q transform with nonstationary Gabor frames},
  author={Velasco, Gino A. and Holighaus, Nicki and D{\"o}rfler, Monika and Grill, Thomas},
  booktitle={Proceedings of the 14th International Conference on Digital Audio Effects (DAFx)},
  year={2011}
}

@article{dupuis2005bayesian,
  title={Bayesian estimation of pulsar parameters from gravitational wave data},
  author={Dupuis, R. J. and Woan, G.},
  journal={Physical Review D},
  volume={72},
  number={10},
  pages={102002},
  year={2005}
}

@article{cornish2010fast,
  title={Fast Bayesian inference for gravitational wave signals},
  author={Cornish, Neil J.},
  journal={Physical Review D},
  volume={82},
  pages={044007},
  year={2010}
}

@book{oppenheim1999signals,
  title={Signals and Systems},
  author={Oppenheim, Alan V. and Willsky, Alan S. and Nawab, S. Hamid},
  year={1999},
  publisher={Prentice Hall}
}

@article{holighaus2013framework,
  title={A framework for invertible, real-time constant-Q transforms},
  author={Holighaus, Nicki and D{\"o}rfler, Monika and Velasco, Gino A. and Grill, Thomas},
  journal={IEEE Transactions on Audio, Speech, and Language Processing},
  volume={21},
  number={4},
  pages={775--785},
  year={2013}
}

@article{waveletinv,
    author = {L. Liu, X. Su  and G. Wang},
    title = {On Inversion of Continuous Wavelet Transform. },
    journal =  {Open Journal of Statistics, 5, 714-720.},
    year = {2015},
    doi = {10.4236/ojs.2015.57071},
}

@article{Gabor1946,
    author = {Gabor, D.},
    title = {Theory of communication. Part 1: The analysis of information},
    journal = {Journal of the Institution of Electrical Engineers - part III: radio and communication engineering},
    volume = {93},
    issue = {26},
    pages = {429--441},
    year = {1946},
}

@article{QTBrown1991,
    author = {Brown, J. C.},
    title = {Calculation of a constant Q spectral transform},
    journal = {The Journal of the Acoustical Society of America},
    volume = {89},
    issue = {1},
    pages = {425--434},
    year = {1991},
}

@book{Mallat2008,
    author = {Mallat, S.},
    title = {A Wavelet Tour of Signal Processing: The Sparse Way},
    publisher = {Academic Press, Inc.},
    year = {2008},
}

@book{Boashash2015,
    author = {Boashash, Boualem},
    title = {Time-frequency signal analysis and processing: a comprehensive reference},
    publisher = {Academic press},
    year = {2015},
}

@article{Klimenko2016,
  title={Method for detection and reconstruction of gravitational wave transients with networks of advanced detectors},
  author={Klimenko, Sergey and others},
  journal={Physical Review D},
  volume={93},
  number={4},
  pages={042004},
  year={2016},
  publisher={APS}
}

@article{Zevin2017,
  title={Gravity Spy: integrating advanced LIGO detector characterization, machine learning, and citizen science},
  author={M. Zevin et al.},
  journal={Classical and Quantum Gravity},
  volume={34},
  number={6},
  pages={064003},
  year={2017},
  publisher={IOP Publishing}
}

@article{Adeli2003,
  title={Analysis of EEG records in an epileptic patient using wavelet transform},
  author={Adeli, H and Ghosh-Dastidar, S and Dadmehr, N},
  journal={Journal of neuroscience methods},
  volume={123},
  number={1},
  pages={69--87},
  year={2003},
  publisher={Elsevier}
}

@article{Addison2005,
  title={Wavelet transforms and the ECG: a review},
  author={Addison, Paul S},
  journal={Physiological measurement},
  volume={26},
  number={5},
  pages={R155},
  year={2005},
  publisher={IOP Publishing}
}

@book{Gencay2001,
  title={An introduction to wavelets and other filtering methods in finance and economics},
  author={Gen{\c{c}}ay, Ramazan and Sel{\c{c}}uk, Faruk and Whitcher, Brandon},
  year={2001},
  publisher={Academic press}
}

@article{Chatterji2004,
    author = {S. Chatterji et al.},
    title = {Multiresolution techniques for the detection of gravitational-wave bursts},
    journal = {Classical and Quantum Gravity},
    volume = {21},
    issue = {20},
    pages = {S1809},
    year = {2004},
}

@article{cWB,
    author = {Marco Drago and Sergey Klimenko and Claudia Lazzaro and Edoardo Milotti and Guenakh Mitselmakher and Valentin Necula and Brendan O’Brian and Giovanni Andrea Prodi and Francesco Salemi and Marek Szczepanczyk and Shubhanshu Tiwari and Vaibhav Tiwari and Gayathri V and Gabriele Vedovato and Igor Yakushin},
    title = {coherent WaveBurst, a pipeline for unmodeled gravitational-wave data analysis},
    journal = {SoftwareX},
    volume = {14},
    pages = {100678},
    year = {2021},
    issn = {2352-7110},
    
}

@article{burst,
  title = {Excess power statistic for detection of burst sources of gravitational radiation},
  author = {Anderson, Warren G. and Brady, Patrick R. and Creighton, Jolien D. E. and Flanagan, \'Eanna \'E.},
  journal = {Phys. Rev. D},
  volume = {63},
  issue = {4},
  pages = {042003},
  numpages = {20},
  year = {2001},
  month = {Jan},
  publisher = {American Physical Society},
  doi = {10.1103/PhysRevD.63.042003},
  url = {https://link.aps.org/doi/10.1103/PhysRevD.63.042003}
}

@article{Bayesinf,
doi = {10.1088/0264-9381/32/13/135012},
url = {https://doi.org/10.1088/0264-9381/32/13/135012},
year = {2015},
month = {jun},
publisher = {IOP Publishing},
volume = {32},
number = {13},
pages = {135012},
author = {Cornish, Neil J and Littenberg, Tyson B},
title = {Bayeswave: Bayesian inference for gravitational wave bursts and instrument glitches},
journal = {Classical and Quantum Gravity},
}

@article{Qptransform,
  title = {Wavelet-based tools to analyze, filter, and reconstruct transient gravitational-wave signals},
  author = {Virtuoso, Andrea and Milotti, Edoardo},
  journal = {Phys. Rev. D},
  volume = {109},
  issue = {10},
  pages = {102010},
  numpages = {13},
  year = {2024},
  month = {May},
  publisher = {American Physical Society},
  doi = {10.1103/PhysRevD.109.102010},
  url = {https://link.aps.org/doi/10.1103/PhysRevD.109.102010}
}

@book{Maggiore2007,
    author = {Maggiore, M.},
    title = {Gravitational Waves. Vol. 1: Theory and Experiments},
    publisher = {Oxford University Press},
    year = {2007},
}

@article{t-SNE,
doi = {10.1088/1361-6382/add3b5},
url = {https://doi.org/10.1088/1361-6382/add3b5},
year = {2025},
month = {may},
publisher = {IOP Publishing},
volume = {42},
number = {10},
pages = {105010},
author = {Ferreira, Tabata Aira and González, Gabriela},
title = {Using t-SNE for characterizing glitches in LIGO detectors},
journal = {Classical and Quantum Gravity}
}

@article{GravitySpy,
    author = {M. Zevin et al.},
    title = {Gravity Spy: integrating advanced LIGO detector characterization, machine learning, and citizen science},
    journal = {Classical and Quantum Gravity},
    volume = {34},
    number = {6},
    pages = {064003},
    year = {2017},
    doi = {10.1088/1361-6382/aa5cea},
}

@article{GlitchFlow2024,
    author = {L. Asprea et al.},
    title = {GlitchFlow, a Digital Twin for transient noise in Gravitational Wave Interferometers},
    journal = {Proceedings of CHEP 2024},
    year = {2024},
}

@article{GSpyNetTree,
    author = {S. Álvarez-López et al.},
    title = {GSpyNetTree: A signal-vs-glitch classifier for gravitational-wave event candidates},
    journal = {Classical and Quantum Gravity},
    volume = {41},
    issue = {8},
    pages = {085007},
    year = {2024},
    doi = {10.1088/1361-6382/ad2194},
}

@article{Gengli,
  title = {Simulating transient noise bursts in LIGO with generative adversarial networks},
  author = {Lopez, Melissa and Boudart, Vincent and Buijsman, Kerwin and Reza, Amit and Caudill, Sarah},
  journal = {Phys. Rev. D},
  volume = {106},
  issue = {2},
  pages = {023027},
  numpages = {17},
  year = {2022},
  month = {Jul},
  publisher = {American Physical Society},
  doi = {10.1103/PhysRevD.106.023027},
  url = {https://link.aps.org/doi/10.1103/PhysRevD.106.023027}
}

@article{GramaxoFreitas2024,
    author = {O. Gramaxo Freitas et al.},
    title = {Assessment of normalizing flows for parameter estimation on time-frequency representations of gravitational-wave data},
    journal = {Machine Learning: Science and Technology},
    volume = {5},
    issue = {1},
    pages = {015036},
    year = {2024},
}

@article{ETDataAnalysis2025,
    author = {R. Abbott et al. (KAGRA Virgo LIGO Scientific)},
    title = {The {S}cience of the {E}instein {T}elescope},
    journal = {arXiv preprint arXiv:2503.12263},
    year = {2025},
}

@article{Macas2022,
    author = {R. Macas et al.},
    title = {Impact of noise transients on low latency gravitational-wave event localization},
    journal = {Phys. Rev. D},
    volume = {105},
    issue = {10},
    pages = {103021},
    year = {2022},
}

@article{lowlatency,
    author = {B.P. Abbott et al.},
    title = {Low-latency Gravitational-wave Alerts for Multimessenger Astronomy during the Second {A}dvanced {LIGO} and {V}irgo {O}bserving {R}un},
    journal = {Astrophys. J.},
    volume = {875},
    issue = {2},
    pages = {161},
    year = {2019},
}

@article{ET,
    author = {M. Punturo et al.},
    title = {The {E}instein {T}elescope: a third-generation gravitational wave observatory},
    journal = {Classical and Quantum Gravity},
    volume = {27},
    pages = {194002},
    year = {2010},
}

@article{CE,
    author = {D. Reitze et al.},
    title = {Cosmic {E}xplorer: The {U}.{S}. Contribution to Gravitational-Wave Astronomy beyond {LIGO}},
    journal = {Bulletin of the American Astronomical Society},
    volume = {51},
    pages = {35},
    year = {2019},
}

@article{overlap,
    author = {P. Relton et al.},
    title = {Addressing the challenges of detecting time-overlapping compact binary coalescences},
    journal = {Phys. Rev. D},
    volume = {106},
    issue = {10},
    pages = {104045},
    year = {2022},
}

@article{Bagnasco2024,
    author = {S. Bagnasco et al.},
    title = {Computing challenges for the {E}instein {T}elescope project},
    journal = {EPJ Web of Conferences},
    volume = {295},
    pages = {04015},
    year = {2024},
}

@article{Cuoco_2021,
doi = {10.1088/2632-2153/abb93a},
url = {https://doi.org/10.1088/2632-2153/abb93a},
year = {2020},
month = {dec},
publisher = {IOP Publishing},
volume = {2},
number = {1},
pages = {011002},
author = {Cuoco, Elena and Powell, Jade and Cavaglià, Marco and Ackley, Kendall and Bejger, Michał and Chatterjee, Chayan and Coughlin, Michael and Coughlin, Scott and Easter, Paul and Essick, Reed and Gabbard, Hunter and Gebhard, Timothy and Ghosh, Shaon and Haegel, Leïla and Iess, Alberto and Keitel, David and Márka, Zsuzsa and Márka, Szabolcs and Morawski, Filip and Nguyen, Tri and Ormiston, Rich and Pürrer, Michael and Razzano, Massimiliano and Staats, Kai and Vajente, Gabriele and Williams, Daniel},
title = {Enhancing gravitational-wave science with machine learning},
journal = {Machine Learning: Science and Technology},
}

@article{Virgotools,
doi = {10.1088/1361-6382/acdf36},
url = {https://doi.org/10.1088/1361-6382/acdf36},
year = {2023},
month = {aug},
publisher = {IOP Publishing},
volume = {40},
number = {18},
pages = {185005},
author = {F Acernese et. al.},
title = {Virgo detector characterization and data quality: tools},
journal = {Classical and Quantum Gravity},
}

@article{Omicron,
title = {Omicron: A tool to characterize transient noise in gravitational-wave detectors},
journal = {SoftwareX},
volume = {12},
pages = {100620},
year = {2020},
issn = {2352-7110},
doi = {https://doi.org/10.1016/j.softx.2020.100620},
url = {https://www.sciencedirect.com/science/article/pii/S2352711020303332},
author = {Florent Robinet and Nicolas Arnaud and Nicolas Leroy and Andrew Lundgren and Duncan Macleod and Jessica McIver}
}

@article{GWpy2021,
    author = {D. M. Macleod et al.},
    title = {GWpy: A Python package for gravitational-wave astrophysics},
    journal = {SoftwareX},
    volume = {13},
    pages = {100657},
    year = {2021},
}

@inproceedings{transformers,
 author = {Vaswani, Ashish and Shazeer, Noam and Parmar, Niki and Uszkoreit, Jakob and Jones, Llion and Gomez, Aidan N and Kaiser, \L ukasz and Polosukhin, Illia},
 booktitle = {Advances in Neural Information Processing Systems},
 editor = {I. Guyon and U. Von Luxburg and S. Bengio and H. Wallach and R. Fergus and S. Vishwanathan and R. Garnett},
 pages = {},
 publisher = {Curran Associates, Inc.},
 title = {Attention is All you Need},
 url = {https://proceedings.neurips.cc/paper_files/paper/2017/file/3f5ee243547dee91fbd053c1c4a845aa-Paper.pdf},
 volume = {30},
 year = {2017}
}

@inproceedings{cnn,
 author = {Krizhevsky, Alex and Sutskever, Ilya and Hinton, Geoffrey E},
 booktitle = {Advances in Neural Information Processing Systems},
 editor = {F. Pereira and C.J. Burges and L. Bottou and K.Q. Weinberger},
 pages = {},
 publisher = {Curran Associates, Inc.},
 title = {ImageNet Classification with Deep Convolutional Neural Networks},
 url = {https://proceedings.neurips.cc/paper_files/paper/2012/file/c399862d3b9d6b76c8436e924a68c45b-Paper.pdf},
 volume = {25},
 year = {2012}
}

@misc{Fessenden1902,
author = {Fessenden, Reginald A.},
title = {Apparatus for Signaling by Electromagnetic Waves},
number = {706,747},
year = {1902},
month = {aug},
day = {12},

}

@book{HaykinCommunicationSystems,
author = {Haykin, Simon S.},
title = {Communication Systems},
publisher = {John Wiley \& Sons},
address = {New York, NY},
edition = {4},
year = {2001},
}

@article{Trovato:2019pr,
  author = "Trovato, Agata",
  title = "{GWOSC: Gravitational Wave Open Science Center}",
  doi = "10.22323/1.357.0082",
  journal = "PoS",
  year = 2019,
  volume = "Asterics2019",
  pages = "082"
}

@article{Samajdar_2021,
   title={Biases in parameter estimation from overlapping gravitational-wave signals in the third-generation detector era},
   volume={104},
   ISSN={2470-0029},
   url={http://dx.doi.org/10.1103/PhysRevD.104.044003},
   DOI={10.1103/physrevd.104.044003},
   number={4},
   journal={Physical Review D},
   publisher={American Physical Society (APS)},
   author={Samajdar, Anuradha and Janquart, Justin and Van Den Broeck, Chris and Dietrich, Tim},
   year={2021},
   month=aug }

@article{Sathyaprakash_2012,
   title={Scientific objectives of {E}instein {T}elescope},
   volume={29},
   ISSN={1361-6382},
   url={http://dx.doi.org/10.1088/0264-9381/29/12/124013},
   DOI={10.1088/0264-9381/29/12/124013},
   number={12},
   journal={Classical and Quantum Gravity},
   publisher={IOP Publishing},
   author={B. Sathyaprakash et al.},
   year={2012},
   month=jun, pages={124013} }

@article{GWTC3,
  author = {R. Abbott et al. (LIGO Scientific Collaboration Virgo Collaboration KAGRA Collaboration)},
  title = {GWTC-3: Compact Binary Coalescences Observed by {LIGO} and {V}irgo during the Second Part of the Third {O}bserving {R}un},
  journal = {Phys. Rev. X},
  volume = {13},
  issue = {4},
  pages = {041039},
  year = {2023}
}

@article{Hannam22,
  author = {M. Hannam  et al.},
  title = {General-relativistic precession in a black-hole binary},
  journal = {Nature},
  volume = {610},
  pages = {652},
  year = {2022}
}

@article{Varma22,
  author = {Varma, V. and et al.},
  title = {Evidence of Large Recoil Velocity from a Black Hole Merger Signal},
  journal = {Phys. Rev. Lett.},
  volume = {128},
  issue = {19},
  pages = {191102},
  year = {2022}
}

@article{Gupte24,
  title = {Evidence for eccentricity in the population of binary black holes observed by LIGO-Virgo-KAGRA},
  author = {Gupte, Nihar and Ramos-Buades, Antoni and Buonanno, Alessandra and Gair, Jonathan and Coleman Miller, M. and Dax, Maximilian and Green, Stephen R. and P\"urrer, Michael and Wildberger, Jonas and Macke, Jakob and Romero-Shaw, Isobel M. and Sch\"olkopf, Bernhard},
  journal = {Phys. Rev. D},
  volume = {112},
  issue = {10},
  pages = {104045},
  numpages = {38},
  year = {2025},
  month = {Nov},
  publisher = {American Physical Society},
  doi = {10.1103/vpyp-nvfp},
  url = {https://link.aps.org/doi/10.1103/vpyp-nvfp}
}

@article{virtuoso_zenodo,
    author = {E. Milotti, A. Virtuoso} ,
    title = {Wavelet-based tools to analyze, filter, and reconstruct transient gravitational-wave signals / Code release (1.0).} ,
    journal = {Zenodo} ,
    year = {2024},
    doi= {https://doi.org/10.5281/zenodo.10649073}
}

@article{Davis22,
  author = {D. Davis et al.},
  title = {Subtracting glitches from gravitational-wave detector data during the third {LIGO}-{V}irgo observing run},
  journal = {Class. Quant. Grav.},
  volume = {39},
  pages = {245013},
  year = {2022}
}

@article{Macas23,
  title = {Revisiting the evidence for precession in {GW}200129 with machine learning noise mitigation},
  author = {Macas, Ronaldas and Lundgren, Andrew and Ashton, Gregory},
  journal = {Phys. Rev. D},
  volume = {109},
  issue = {6},
  pages = {062006},
  numpages = {8},
  year = {2024},
  month = {Mar},
  publisher = {American Physical Society},
  doi = {10.1103/PhysRevD.109.062006},
  url = {https://link.aps.org/doi/10.1103/PhysRevD.109.062006}
}

@article{Payne22,
  author = {E. Payne et al.},
  title = {Curious case of {GW}200129: Interplay between spin-precession inference and data-quality issues},
  journal = {Phys. Rev. D},
  volume = {106},
  pages = {104017},
  year = {2022}
}

@article{ml4gw,  year = {2025}, publisher = {The Open Journal}, volume = {10}, number = {114}, pages = {8836}, author = {W. Benoit  E. Marx D. Chatterjee R. Kumar and A. Gunny}, title = {ml4gw: PyTorch utilities for training neural networks in gravitational wave physics applications}, journal = {Journal of Open Source Software} }

@article{Ashton19,
  author = {G. Ashton et al.},
  title = {{BILBY: A user-friendly Bayesian inference library for gravitational-wave astronomy}},
  journal = {Astrophys. J. Suppl.},
  volume = {241},
  number = {2},
  pages = {27},
  year = {2019},
  doi = {10.3847/1538-4365/ab06fc}
}

@article{Driggers19,
  author = {J. C. Driggers et al.},
  title = {{Improving astrophysical parameter estimation via offline noise subtraction for Advanced LIGO}},
  journal = {Phys. Rev. D},
  volume = {99},
  number = {4},
  pages = {042001},
  year = {2019},
  doi = {10.1103/PhysRevD.99.042001}
}

@article{Cornish15,
  author = {Cornish, Neil J. and Littenberg, Tyson B.},
  title = {{BayesWave: Bayesian Inference for Gravitational Wave Bursts and Instrument Glitches}},
  journal = {Class. Quant. Grav.},
  volume = {32},
  number = {13},
  pages = {135012},
  year = {2015},
  doi = {10.1088/0264-9381/32/13/135012}
}

@article{Littenberg15,
  author = {Littenberg, Tyson B. and Cornish, Neil J.},
  title = {{Bayesian inference for spectral estimation of gravitational wave detector noise}},
  journal = {Phys. Rev. D},
  volume = {91},
  number = {8},
  pages = {084034},
  year = {2015},
  doi = {10.1103/PhysRevD.91.084034}
}

@article{Kingsbury2001,
  title={Complex wavelets for shift invariant analysis and filtering of signals},
  author={Kingsbury, Nick},
  journal={Applied and computational harmonic analysis},
  volume={10},
  number={3},
  pages={234--253},
  year={2001},
  publisher={Elsevier}
}

@misc{UnoaccasoRepo,
  author = {Felicetti, Riccardo},
  title = {TimeSeries Project},
  year = {2024},
  publisher = {GitHub},
  journal = {GitHub repository},
  howpublished = {\url{https://github.com/Unoaccaso/timeseries}},
  note = {Accessed: 2025-11-03}
}

@article{PhysRevD.108.124037,
  title = {Next generation of accurate and efficient multipolar precessing-spin effective-one-body waveforms for binary black holes},
  author = {Ramos-Buades, Antoni and Buonanno, Alessandra and Estell\'es, H\'ector and Khalil, Mohammed and Mihaylov, Deyan P. and Ossokine, Serguei and Pompili, Lorenzo and Shiferaw, Mahlet},
  journal = {Phys. Rev. D},
  volume = {108},
  issue = {12},
  pages = {124037},
  numpages = {34},
  year = {2023},
  month = {Dec},
  publisher = {American Physical Society},
  doi = {10.1103/PhysRevD.108.124037},
  url = {https://link.aps.org/doi/10.1103/PhysRevD.108.124037}
}

@book{grochenig2001foundations,
  title={Foundations of Time-Frequency Analysis},
  author={Gr{\"o}chenig, Karlheinz},
  year={2001},
  publisher={Birkh{\"a}user}
}

@patent{Bell1876,
  author = {Alexander Graham Bell},
  title = {Improvement in Telegraphy},
  number = {U.S. Patent 174,465},
  year = {1876},
  month = {March}
}

@patent{Landell1901,
  author = {Roberto Landell de Moura},
  title = {Phonetic transmission at a distance, with or without wire, through space, Earth and water},
  number = {Brazilian Patent 3,279},
  year = {1901},
  month = {March}
}

@article{Armstrong1914,
  author = {Edwin H. Armstrong},
  title = {Some Recent Developments in the Audion Receiver},
  journal = {Proceedings of the Institute of Radio Engineers},
  volume = {2},
  number = {3},
  pages = {215--247},
  year = {1914},
  publisher = {IEEE}
}

@article{cWB-XP,
  title = {Coherent WaveBurst with eXtended Pixels for Gravitational-Wave Burst Searches},
  author = {Mishra, T. and Tiwari, V. and Klimenko, S. and Vedovato, G. and Mitselmakher, G.},
  journal = {Physical Review D},
  volume = {104},
  issue = {2},
  pages = {023014},
  year = {2021},
  doi = {10.1103/PhysRevD.104.023014},
  publisher = {American Physical Society}
}

@article{scatteringtransform,
  author={J.Bruna S. Mallat},
  journal={IEEE Transactions on Pattern Analysis and Machine Intelligence}, 
  title={Invariant Scattering Convolution Networks}, 
  year={2013},
  volume={35},
  number={8},
  pages={1872-1886},
  doi={10.1109/TPAMI.2012.230}}

\end{document}